\documentclass[12pt]{article}

\usepackage{scicite}

\usepackage{times}
\usepackage{amsmath, amssymb}
\usepackage{url}
\usepackage{graphicx}
\usepackage{subcaption}
\usepackage{caption}
\usepackage{threeparttable}
\usepackage{multibib}
\usepackage{svg}
\usepackage[amssymb]{SIunits}
\usepackage{ulem}

\usepackage{soul}
\usepackage{verbatim}
\usepackage[left]{lineno}

\renewcommand{\theequation}{{S\arabic{equation}}}

\def\X-#1{X\nobreakdash-#1}

\newcites{main}{References}
\newcites{methods}{References for Methods}

\newenvironment{sciabstract}{%
\begin{quote} \bf}
{\end{quote}}

\title{Observations of a black hole X-ray binary indicate formation of a magnetically arrested disk} 

\author
{Bei\ You$^{1,\ast}$, Xinwu\ Cao$^{2,\ast}$, Zhen Yan$^{3,\ast}$, Jean-Marie Hameury$^{4}$,\\ Bozena Czerny$^{5}$, Yue Wu$^{1,6,7}$, Tianyu Xia$^{1,8,9}$, Marek Sikora$^{10}$,\\ Shuang-Nan Zhang$^{11,12}$, Pu Du$^{11}$, Piotr T. Zycki $^{10}$}

\date{}

\begin{document} 


\baselineskip 24pt


\maketitle

\noindent
1:  Department of Astronomy, School of Physics and Technology, Wuhan University, Wuhan 430072, China\\
2:  Institute for Astronomy, School of Physics, Zhejiang University, Hangzhou 310058, China\\
3:  Shanghai Astronomical Observatory, Chinese Academy of Sciences (CAS), Shanghai 200030, China\\
4:  Observatoire Astronomique de Strasbourg, Université de Strasbourg and Centre National de la Recherche Scientifique, Unite Mixte de Recherche 7550, 67000 Strasbourg, France\\
5:  Center for Theoretical Physics, Polish Academy of Sciences, Warsaw 02-668, Poland\\
6:  School of Astronomy and Space Science, Nanjing University, Nanjing 210023, China 
\\
7:  Key laboratory of Modern Astronomy and Astrophysics (Nanjing University), Ministry of Education, Nanjing 210023, China\\
8:  CAS Key laboratory for Research in Galaxies and Cosmology, Department of Astronomy, University of Science and Technology of China, Hefei 230026, China
\\
9:  School of Astronomy and Space Sciences, University of Science and Technology of China, Hefei 230026, China\\
10:  Nicolaus Copernicus Astronomical Center, Polish Academy of Sciences, Warsaw 00-716, Poland\\
11:  Key Laboratory for Particle Astrophysics, Institute of High Energy Physics, Chinese Academy of Sciences, Beijing 100049, China\\
12:  University of Chinese Academy of Sciences, Chinese Academy of Sciences, Beijing 100049, China\\

Corresponding authors \\
E-mail: youbei@whu.edu.cn, xwcao@zju.edu.cn, zyan@shao.ac.cn


\begin{sciabstract}
Accretion of material onto a black hole drags any magnetic fields present inwards, increasing their strength. Theory predicts that sufficiently strong magnetic fields can halt the accretion flow, producing a magnetically arrested disk (MAD). We analyze archival multi-wavelength observations of an outburst from the black hole X-ray binary MAXI J1820+070 in 2018. The radio and optical fluxes are delayed by about 8 and 17 days respectively, compared to the X-ray flux. We interpret this as evidence for the formation of a MAD. In this scenario, the magnetic field is amplified by an expanding corona, forming a MAD around the time of the radio peak. The optical delay is then due to thermal viscous instability in the outer disk. 
\end{sciabstract}



A black hole X-ray binary (BHXRB) is a binary system that consists of a black hole (BH) and a normal star, which emits large amounts of radiation in the X-ray band. 
The gas of the star drawn by the strong gravity flows towards the BH and forms an accretion disk (namely accretion flow). In the disk, matter moves inwards under the effect of viscosity. 
Relativistic jets can be launched near a BH due to magnetic fields carried by the accretion flow in the vicinity of the BH \cite{blandford1977,blandford1982}. Numerical simulations have shown that any large-scale magnetic field present in the accreting material can be dragged inwards through the accretion flow, increasing its strength. If the radial magnetic force becomes sufficiently high that it equals the gravitational force of the BH, the accretion flow halts and a magnetically arrested disk (MAD) is formed \cite{igumenshchev2003,narayan2003}. 

Sufficiently strong magnetic field strengths to produce a MAD have been observationally inferred in accretion systems with supermassive BHs \cite{zamaninasab2014,eht2021,eht2022,yuan2022}. MADs are also predicted in BHXRBs, though there is little observational evidence for these \cite{thomas2022}. The formation process of a MAD has not been observed in either type of BH (supermassive or stellar-mass). Accretion flows onto stellar-mass BHs can evolve much faster than those around supermassive BHs. For example, outbursts in BHXRBs last for months to years, potentially allowing multi-wavelength observations to probe the formation of a MAD.

The outbursts of BHXRB are well explained by the disk-instability model\cite{lasota2001}. In quiescence, mass accumulates in a cold, geometrically thin accretion disk (i.e. thin disk) until a thermal-viscous instability is triggered when, somewhere in the disk, the temperature reaches the hydrogen ionization temperature. 
An outburst then starts and the disk empties. 
A BHXRB usually experiences two very different spectral states during an outburst, i.e. a hard state and a soft state, with the X-ray emission dominated by either hard ($>$10 keV) or soft ($<$ 10 keV) X-rays. The timing, and radio emission properties also differ between the two states \cite{remillard2006,fender2014}. The period between the two states is called state transition. 

In the hard state, the hot gas near the BH forms the corona which is modeled as an advection-dominated accretion flow (ADAF), since a significant fraction of the viscously released gravitational energy is advected with the flow and enters the event horizon \cite{narayan1994}. Hard states are observed during the rise and decay of an outburst, and are referred to as the rising and decaying hard state in the following. The soft-to-hard state transition and the decaying hard state are accompanied by a hard X-ray flare. 
The flare is also detected in the optical, infrared, and radio bands \cite{russell2010, corbel2013}, which indicates the formation of a jet. However, the detailed mechanisms and processes are still unclear. Hard X-rays during the flare are usually thought to be emitted by inverse Compton scattering in the jet and/or in the ADAF \cite{remillard2006}.

\subsection*{An outburst of MAXI J1820+070 in 2018} 
MAXI J1820+070 (also known as ASASSN-18ey) is an X-ray binary consisting of a low-mass star orbiting a stellar-mass BH (BH mass $M_{\rm BH}\simeq 8.5M_{\odot}$), with an orbital period of 16.5 hours \cite{torres2020}. It was discovered during an X-ray outburst that occurred on 2018 March 11 \cite{kawamuro2018}. It is located at a distance of $2.96 \pm 0.33$ kiloparsecs (kpc) \cite{atri2020}. The outburst was observed by several X-ray missions \cite{you2021,ma2021,buisson2019, kara2019}. We analyze archival observations by the \textit{Insight-Hard X-ray Modulation Telescope} [\textit{Insight}-HXMT,\cite{zhang2020}]. The \textit{Insight}-HXMT data cover a broad energy band of 1--250 keV, from 2018 March 14 to 2018 October 21 corresponding to modified Julian date (MJD) 58191-58412 \cite{you2021,ma2021}. MAXI J1820+070 was also an optically bright transient \cite{littlefield2018}, so was monitored by the American Association of Variable Star Observers (AAVSO) which produced a light curve (brightness as a function of time) from 2018 March 12 to 2018 December 21 (MJD 58188-59184) \cite{patterson2018}.  Radio emission from MAXI J1820+070 was also detected by multiple telescopes, indicating the presence of a jet. We use the radio data from Arcminute Microkelvin Imager Large Array (AMI-LA) \cite{bright2020}. These datasets provide high-cadence monitoring of the 2018 outburst of MAXI J1820+070 over a broad range of wavelengths. The proximity of the source and its low interstellar extinction of E(B-V)=0.20--0.26 \cite{mikolajewska2022} made it sufficiently bright to investigate the accretion process of the magnetized flow near a BH. 

\begin{figure}
\centering
\includegraphics[width=\textwidth]{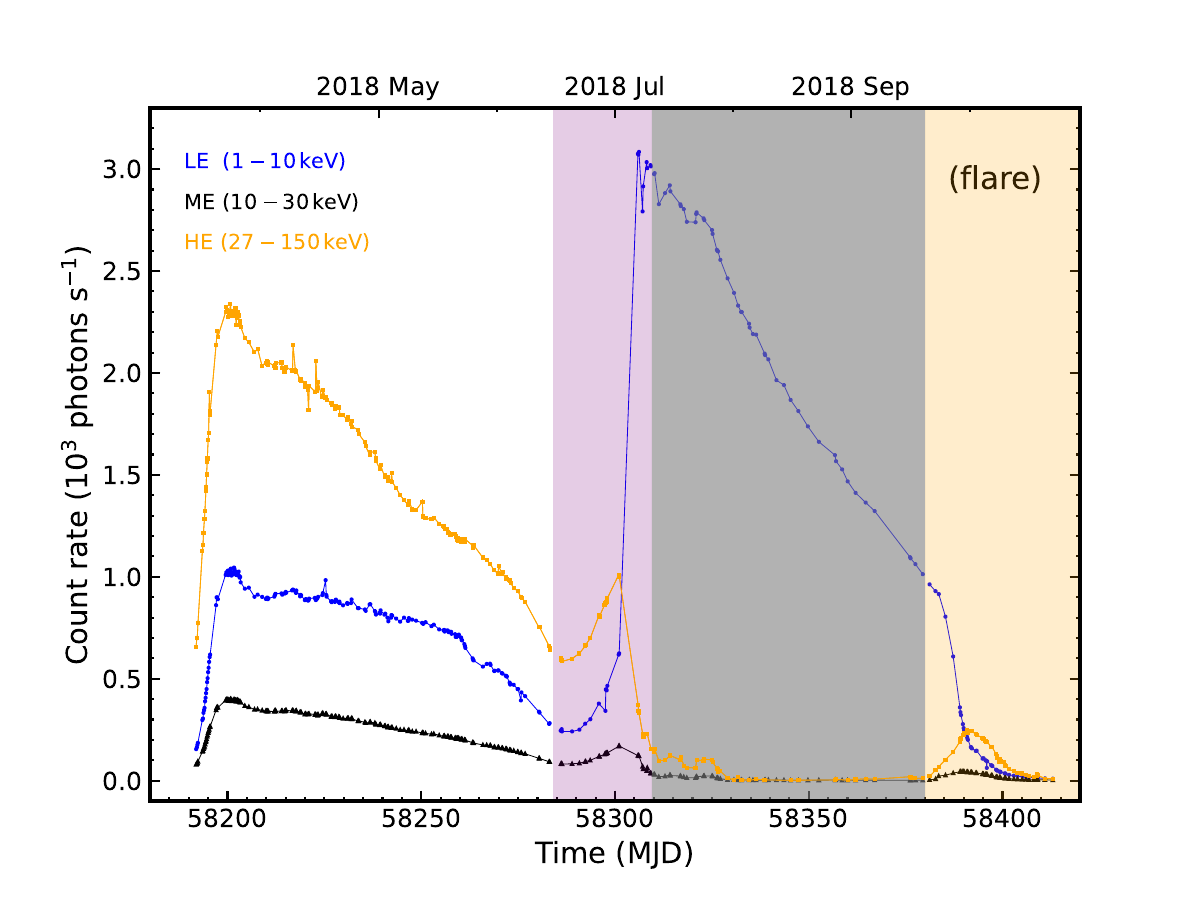}
\caption{
 \textit{Insight}-HXMT light curves of MAXI J1820+070 during the 2018 outburst. The orange, black and skyblue points, correspond to the light curves by HE (27--150 keV), ME (10--30 keV), and LE (1--10 keV), respectively. The unshaded region, the light purple region, and the light grey region indicate the periods in which the source was in the rising hard state, the hard-to-soft transition state, and the soft state, respectively. The flare, covering the soft-to-hard transition state and the decaying hard state, is marked by the light orange region. The error bars, which are too small to be seen for most data, correspond to a 1-$\sigma$ confidence interval.
 }
\label{hxmt_lc}
\end{figure}

Fig. \ref{hxmt_lc} shows the \textit{Insight}-HXMT light curves from the low energy (LE, 1--10 keV), medium energy (ME, 10--30 keV), and high energy (HE, 27--150 keV) instruments, between MJD 58191 and 58412. We focus our analysis on the radio, optical, and X-ray observations of MAXI J1820+070 during the hard X-ray flare including the soft-to-hard state transition and decaying hard state periods (MJD 58380--58412). 

\subsection*{Lags between flares of different wavelengths} 
Fig. \ref{multi_lc}A shows the X-ray luminosity of MAXI J1820+070 during the flare, which we determined from the \textit{Insight}-HXMT data \cite{methods}. 
The luminosities of the disk and Comptonization components derived from X-ray spectral fitting are also shown. 
Fig. \ref{multi_lc}B shows the radio brightness at 15.5 GHz derived from AMI-LA observations, which exhibits a radio flare. The optical ($V$-band) data from AAVSO are shown in Fig. \ref{multi_lc}C, which also exhibits a flare superimposed on the exponential decay of the underlying outburst. 

We performed an interpolated cross-correlation function (ICCF) analysis \cite{methods} to determine the time lag between the radio and X-ray light curves, finding a radio lag of $8.06^{+0.88}_{-0.56}$ days (Fig. \ref{58380_r_v_x}A). 
ICCF analysis comparing the radio and the $V$-band shows an optical lag of $8.61^{+1.06}_{-0.83}$ days with respect to the radio, indicating negligible contribution from the jet to the $V$-band flux (see Supplementary Text). 
ICCF analysis comparing the X-ray and the $V$-band shows an optical lag of $17.61^{+0.50}_{-0.44}$ days with respect to the X-ray (Fig. \ref{58380_r_v_x}B).
This lag is far longer than the time taken for light to travel across the binary, which is several seconds for an orbital period of 16.5 hours, so cannot be due to reprocessed X-ray illumination of the outer disk or the companion star. The uncertainties of the ICCF analysis above are quoted at the 1-$\sigma$ level. Besides the cross-correlation analysis of those light curves shown above, the correlations between the X-ray and radio/{\it V}-band luminosities are also used to study the origins of the emission \cite{methods}.

\section*{Interpretation of the radio time lag} 
The hard X-ray emission during this flare comes from soft X-ray photons inverse Compton scattered by hot electrons (i.e., Comptonization) within the jet or the ADAF. Given that (i) the radio emission originates from the jet \cite{bright2020}, and (ii) the radio flare lags the X-ray by $\sim$ 8 days,
we rule out the possibility that X-rays are emitted by the jet, so suggest that the hard X-rays were most likely emitted by the ADAF. 

In the hard state of BHXRB, a geometrically thick ADAF is thought to locate close to BH, extending outwards and connecting to a thin disk with a truncated inner radius $R_{\rm tr}$ (equals to the ADAF radial extent)\cite{yuan2014,liu2022}.  
We assume the truncated radius was $R_{\rm tr} = R_{\rm ISCO}$ at $t_0$ = MJD 58381 when the soft state ends, where $R_{\rm ISCO}$ is the innermost stable circular orbit (ISCO). Based on the derived disk luminosity which is estimated from the spectral fitting \cite{methods}, we find that the truncation radius increases to $R_{\rm tr} = 6.79 R_{\rm ISCO}$ when the hard X-ray emission peaks at $t_1$ = MJD 58389, and further increases to $R_{\rm tr} = 32.83 R_{\rm ISCO}$ when the radio emission peaks on $t_2$ = MJD 58397 \cite{methods}.
$R_{\rm tr}$ therefore increases with decreasing mass accretion rate. In this case, the fraction of the total gravitational power dissipated in the whole ADAF increases with $R_{\rm tr}$, while the power in the very inner part of ADAF, where the hard X-ray mainly comes from, decreases with decreasing accretion rate [\cite{methods}, Equation S5].
These two mechanisms compete with each other, so a peak in the hard X-ray emission occurs when the latter mechanism becomes dominant, around $t_1$ = MJD 58389. 

Previous studies have suggested that a weak and coherent external magnetic field can be dragged inwards and amplified by an ADAF, due to its high radial velocity \cite{narayan2012,begelman2022}. The external magnetic field is expected to be more strongly amplified by larger ADAFs \cite{cao2011}. 
We therefore calculated the advection and diffusion of magnetic fields in the ADAF of MAXI J1820+070 \cite{methods}. We find the magnetic field at $R_{\rm ISCO}$ is $B \simeq 2\times 10^6 ~ \rm G$ at $t_0$ = MJD 58381 (the onset of the flare), $B \simeq 4.5\times 10^7 ~ \rm G$ at $t_1$ = MJD 58389 (the peak of the hard X-ray emission), saturates at $B \simeq 7\times 10^7 ~ \rm G$ around $t_2$ = MJD 58397, and then declines. 
The jet power increases with the magnetic field strength near the BH, so its radio emission also increases until $t_2$ = MJD 58397, explaining the observed delay of $\sim 8$ days with respect to the X-ray emission. 
The jet power at the radio peak is estimated as $10^{37}$ erg s$^{-1}$, consistent with the theoretical prediction from the magnetic field of $7\times 10^7 ~ \rm Gauss$ \cite{methods}, which supports our interpretation that the magnetic field is transported and amplified by the expanding ADAF.

\subsection*{Formation of a magnetically arrested disk}
In our interpretation, the increasingly strong magnetic field influences the subsequent accretion flow, because the magnetic pressure acts against the gravity of the BH. According to our calculations of the radius-dependent magnetic fields within the ADAF \cite{methods}, the BH gravity still dominates over magnetic pressure at the hard X-ray peak ($t_1$ = MJD 58389). After the hard X-ray peak, the magnetic field inside the expanding ADAF continues to be amplified, eventually becoming dominant at the inner edge, so a MAD forms at the time of the radio peak ($t_2$ = MJD 58397) when the magnetic pressure becomes equal to gravity force [\cite{methods}, Fig. \ref{b_t}].
 
We conclude that the increase of the truncated radius during the flare in MAXI J1820+70 leads to the observed 8-day radio lag relative to the hard X-ray emission. The MAD is then gradually established in the ADAF as it continues to extend radially. Fig. \ref{schematic_MAD} illustrates the physical process associated with the X-ray/radio flare and the formation of the MAD in our proposed scenario. An animated version is provided in Movie S1. 

\subsection*{Interpretation of the optical time lag} 
We consider whether the long delay time of the optical flare can be produced by a viscously heated thin disk.
In this scenario, the X-ray flare heats up the outer thin disk, reviving the thermal-viscous instability in the thin disk, which produces a delayed optical emission.
We use a disk instability code \cite{hameury1998} to determine the time-dependent evolution of the thin disk. The results are shown in Fig. \ref{flare}. 
We find a small optical flare is produced, starting around MJD 58390 but fading away after a few days, which is much shorter than the observed lag of $\sim$17 days. Therefore, the standard disk instability model (DIM) cannot explain this optical lag. 

A thin disk is expected to launch disk winds, which has previously been studied using observations and simulations \cite{miller2006,you2016}. Previous work has found that the disk wind is present throughout the outburst of MAXI J1820+070 \cite{sanchez2020}.
We therefore postulate that the disk wind is launched during the flare. Disk winds affect the profile of the outburst light curves of BHXRBs \cite{tetarenko2018, dubus2019}, because the disk wind removes angular momentum from the disk, which is equivalent to increasing the viscosity. As a result, a brighter and more delayed optical peak is produced comparing to the case of no disk wind. Using a simple parametrization of the disk wind, 
we performed additional simulations of the DIM \cite{methods}. We find that the resulting $V$-band flare is consistent with the observed light curve, and lags the hard X-rays by more than 15 days (Fig. \ref{flare}).




\clearpage

\bibliographystyle{Science}
\bibliography{reference.bib}

\begin{thebibliography}{10}

\bibitem{blandford1977}
R.~D. {Blandford}, R.~L. {Znajek}, {\it \mnras\/} {\bf 179}, 433 (1977).

\bibitem{blandford1982}
R.~D. {Blandford}, D.~G. {Payne}, {\it \mnras\/} {\bf 199}, 883 (1982).

\bibitem{igumenshchev2003}
I.~V. {Igumenshchev}, R.~{Narayan}, M.~A. {Abramowicz}, {\it \apj\/} {\bf 592},
  1042 (2003).

\bibitem{narayan2003}
R.~{Narayan}, I.~V. {Igumenshchev}, M.~A. {Abramowicz}, {\it \pasj\/} {\bf 55},
  L69 (2003).

\bibitem{zamaninasab2014}
M.~{Zamaninasab}, E.~{Clausen-Brown}, T.~{Savolainen}, A.~{Tchekhovskoy}, {\it
  \nat\/} {\bf 510}, 126 (2014).

\bibitem{eht2021}
{Event Horizon Telescope Collaboration}, {\it et~al.\/}, {\it \apjl\/} {\bf
  910}, L13 (2021).

\bibitem{eht2022}
{Event Horizon Telescope Collaboration}, {\it et~al.\/}, {\it \apjl\/} {\bf
  930}, L16 (2022).

\bibitem{yuan2022}
F.~{Yuan}, H.~{Wang}, H.~{Yang}, {\it arXiv e-prints\/} p. arXiv:2201.00512
  (2022).

\bibitem{thomas2022}
J.~K. {Thomas}, {\it et~al.\/}, {\it \mnras\/} {\bf 509}, 1062 (2022).

\bibitem{lasota2001}
J.-P. {Lasota}, {\it \nar\/} {\bf 45}, 449 (2001).

\bibitem{remillard2006}
R.~A. {Remillard}, J.~E. {McClintock}, {\it \araa\/} {\bf 44}, 49 (2006).

\bibitem{fender2014}
R.~{Fender}, E.~{Gallo}, {\it \ssr\/} {\bf 183}, 323 (2014).

\bibitem{narayan1994}
R.~{Narayan}, I.~{Yi}, {\it \apjl\/} {\bf 428}, L13 (1994).

\bibitem{russell2010}
D.~M. {Russell}, D.~{Maitra}, R.~J.~H. {Dunn}, S.~{Markoff}, {\it MNRAS\/} {\bf
  405}, 1759 (2010).

\bibitem{corbel2013}
S.~{Corbel}, {\it et~al.\/}, {\it \mnras\/} {\bf 428}, 2500 (2013).

\bibitem{torres2020}
M.~A.~P. {Torres}, {\it et~al.\/}, {\it \apjl\/} {\bf 893}, L37 (2020).

\bibitem{kawamuro2018}
T.~{Kawamuro}, {\it et~al.\/}, {\it The Astronomer's Telegram\/} {\bf 11399}, 1
  (2018).

\bibitem{atri2020}
P.~{Atri}, {\it et~al.\/}, {\it \mnras\/} {\bf 493}, L81 (2020).

\bibitem{you2021}
B.~{You}, {\it et~al.\/}, {\it Nature Communications\/} {\bf 12}, 1025 (2021).

\bibitem{ma2021}
X.~{Ma}, {\it et~al.\/}, {\it Nature Astronomy\/} {\bf 5}, 94 (2021).

\bibitem{buisson2019}
D.~J.~K. {Buisson}, {\it et~al.\/}, {\it \mnras\/} {\bf 490}, 1350 (2019).

\bibitem{kara2019}
E.~{Kara}, {\it et~al.\/}, {\it Nature\/} {\bf 565}, 198 (2019).

\bibitem{zhang2020}
S.-N. {Zhang}, {\it et~al.\/}, {\it Science China Physics, Mechanics, and
  Astronomy\/} {\bf 63}, 249502 (2020).

\bibitem{littlefield2018}
C.~{Littlefield}, {\it The Astronomer's Telegram\/} {\bf 11421}, 1 (2018).

\bibitem{patterson2018}
J.~{Patterson}, {\it et~al.\/}, {\it The Astronomer's Telegram\/} {\bf 11756},
  1 (2018).

\bibitem{bright2020}
J.~S. {Bright}, {\it et~al.\/}, {\it Nature Astronomy\/} {\bf 4}, 697 (2020).

\bibitem{mikolajewska2022}
J.~{Miko{\l}ajewska}, A.~A. {Zdziarski}, J.~{Zi{\'o}{\l}kowski}, M.~A.~P.
  {Torres}, J.~{Casares}, {\it \apj\/} {\bf 930}, 9 (2022).

\bibitem{methods}
{\it Materials and methods are available as supplementary materials\/} .

\bibitem{yuan2014}
F.~{Yuan}, R.~{Narayan}, {\it \araa\/} {\bf 52}, 529 (2014).

\bibitem{liu2022}
B.~F. {Liu}, E.~{Qiao}, {\it iScience\/} {\bf 25}, 103544 (2022).

\bibitem{narayan2012}
R.~{Narayan}, A.~{S{\"A} dowski}, R.~F. {Penna}, A.~K. {Kulkarni}, {\it
  \mnras\/} {\bf 426}, 3241 (2012).

\bibitem{begelman2022}
M.~C. {Begelman}, N.~{Scepi}, J.~{Dexter}, {\it \mnras\/} {\bf 511}, 2040
  (2022).

\bibitem{cao2011}
X.~{Cao}, {\it \apj\/} {\bf 737}, 94 (2011).

\bibitem{hameury1998}
J.-M. {Hameury}, K.~{Menou}, G.~{Dubus}, J.-P. {Lasota}, J.-M. {Hure}, {\it
  \mnras\/} {\bf 298}, 1048 (1998).

\bibitem{miller2006}
J.~M. {Miller}, {\it et~al.\/}, {\it Nature\/} {\bf 441}, 953 (2006).

\bibitem{you2016}
B.~{You}, {\it et~al.\/}, {\it \apj\/} {\bf 821}, 104 (2016).

\bibitem{sanchez2020}
J.~{S{\'a}nchez-Sierras}, T.~{Mu{\~n}oz-Darias}, {\it \aaa\/} {\bf 640}, L3
  (2020).

\bibitem{tetarenko2018}
B.~E. {Tetarenko}, J.~P. {Lasota}, C.~O. {Heinke}, G.~{Dubus}, G.~R.
  {Sivakoff}, {\it Nature\/} {\bf 554}, 69 (2018).

\bibitem{dubus2019}
G.~{Dubus}, C.~{Done}, B.~E. {Tetarenko}, J.-M. {Hameury}, {\it \aaa\/} {\bf
  632}, A40 (2019).

\bibitem{mad2023}
B.~You, X.~Cao, Z.~Yan, {Observations of a black hole X-ray binary indicate
  formation of a magnetically arrested disk} (2023).
  \url{doi.org/10.5281/zenodo.8080948}.

\bibitem{arnaud1996}
K.~A. {Arnaud}, {\it Astronomical Data Analysis Software and Systems V\/},
  G.~H. {Jacoby}, J.~{Barnes}, eds. (1996), vol. 101 of {\it Astronomical
  Society of the Pacific Conference Series\/}, p.~17.

\bibitem{frei1994}
Z.~{Frei}, J.~E. {Gunn}, {\it \aj\/} {\bf 108}, 1476 (1994).

\bibitem{oke1974}
J.~B. {Oke}, {\it \apjs\/} {\bf 27}, 21 (1974).

\bibitem{kaspi2000}
S.~{Kaspi}, {\it et~al.\/}, {\it \apj\/} {\bf 533}, 631 (2000).

\bibitem{bentz2009}
M.~C. {Bentz}, {\it et~al.\/}, {\it \apj\/} {\bf 705}, 199 (2009).

\bibitem{du2014}
P.~{Du}, {\it et~al.\/}, {\it \apj\/} {\bf 782}, 45 (2014).

\bibitem{king1998}
A.~R. {King}, H.~{Ritter}, {\it \mnras\/} {\bf 293}, L42 (1998).

\bibitem{blandford1999}
R.~D. {Blandford}, M.~C. {Begelman}, {\it \mnras\/} {\bf 303}, L1 (1999).

\bibitem{xie2019}
F.-G. {Xie}, A.~A. {Zdziarski}, {\it \apj\/} {\bf 887}, 167 (2019).

\bibitem{cao2016}
X.~{Cao}, {\it \apj\/} {\bf 817}, 71 (2016).

\bibitem{shakura1973}
N.~I. {Shakura}, R.~A. {Sunyaev}, {\it \aaa\/} {\bf 500}, 33 (1973).

\bibitem{bisnovatyi1974}
G.~S. {Bisnovatyi-Kogan}, A.~A. {Ruzmaikin}, {\it Astrophys. Space Sci.\/} {\bf
  28}, 45 (1974).

\bibitem{vanballegooijen1989}
A.~A. {van Ballegooijen}, {\it Accretion Disks and Magnetic Fields in
  Astrophysics\/}, G.~{Belvedere}, ed. (1989), vol. 156 of {\it Astrophysics
  and Space Science Library\/}, p.~99.

\bibitem{lubow1994}
S.~H. {Lubow}, J.~C.~B. {Papaloizou}, J.~E. {Pringle}, {\it \mnras\/} {\bf
  268}, 1010 (1994).

\bibitem{cao2020}
X.~{Cao}, A.~A. {Zdziarski}, {\it \mnras\/} {\bf 492}, 223 (2020).

\bibitem{donati2009}
J.~F. {Donati}, J.~D. {Landstreet}, {\it \araa\/} {\bf 47}, 333 (2009).

\bibitem{cao2019}
X.~{Cao}, D.~{Lai}, {\it \mnras\/} {\bf 485}, 1916 (2019).

\bibitem{tout1996}
C.~A. {Tout}, J.~E. {Pringle}, {\it \mnras\/} {\bf 281}, 219 (1996).

\bibitem{beckwith2009}
K.~{Beckwith}, J.~F. {Hawley}, J.~H. {Krolik}, {\it \apj\/} {\bf 707}, 428
  (2009).

\bibitem{liska2020}
M.~{Liska}, A.~{Tchekhovskoy}, E.~{Quataert}, {\it \mnras\/} {\bf 494}, 3656
  (2020).

\bibitem{cao2021}
X.~{Cao}, B.~{You}, Z.~{Yan}, {\it \aaa\/} {\bf 654}, A81 (2021).

\bibitem{king2007}
A.~R. {King}, J.~E. {Pringle}, M.~{Livio}, {\it \mnras\/} {\bf 376}, 1740
  (2007).

\bibitem{parker1979}
E.~N. {Parker}, {\it {Cosmical magnetic fields. Their origin and their
  activity}\/} ({Oxford: Clarendon Press}, 1979).

\bibitem{fromang2009}
S.~{Fromang}, J.~M. {Stone}, {\it \aaa\/} {\bf 507}, 19 (2009).

\bibitem{guan2009}
X.~{Guan}, C.~F. {Gammie}, {\it \apj\/} {\bf 697}, 1901 (2009).

\bibitem{lesur2009}
G.~{Lesur}, P.~Y. {Longaretti}, {\it \aaa\/} {\bf 504}, 309 (2009).

\bibitem{igumenshchev2002}
I.~V. {Igumenshchev}, R.~{Narayan}, {\it \apj\/} {\bf 566}, 137 (2002).

\bibitem{kawamura2022}
T.~{Kawamura}, M.~{Axelsson}, C.~{Done}, T.~{Takahashi}, {\it \mnras\/} {\bf
  511}, 536 (2022).

\bibitem{narayan2022}
R.~{Narayan}, A.~{Chael}, K.~{Chatterjee}, A.~{Ricarte}, B.~{Curd}, {\it
  \mnras\/} {\bf 511}, 3795 (2022).

\bibitem{tetarenko2021}
B.~E. {Tetarenko}, {\it et~al.\/}, {\it \mnras\/} {\bf 501}, 3406 (2021).

\bibitem{zdziarski2022}
A.~A. {Zdziarski}, A.~J. {Tetarenko}, M.~{Sikora}, {\it \apj\/} {\bf 925}, 189
  (2022).

\bibitem{falcke1996}
H.~{Falcke}, P.~L. {Biermann}, {\it \aaa\/} {\bf 308}, 321 (1996).

\bibitem{coriat2009}
M.~{Coriat}, {\it et~al.\/}, {\it \mnras\/} {\bf 400}, 123 (2009).

\bibitem{livio1999}
M.~{Livio}, G.~I. {Ogilvie}, J.~E. {Pringle}, {\it \apj\/} {\bf 512}, 100
  (1999).

\bibitem{smak1984}
J.~{Smak}, {\it Acta Astron.\/} {\bf 34}, 161 (1984).

\bibitem{hirose2014}
S.~{Hirose}, O.~{Blaes}, J.~H. {Krolik}, M.~S.~B. {Coleman}, T.~{Sano}, {\it
  \apj\/} {\bf 787}, 1 (2014).

\bibitem{dubus2001}
G.~{Dubus}, J.~M. {Hameury}, J.~P. {Lasota}, {\it \aaa\/} {\bf 373}, 251
  (2001).

\bibitem{menou2000}
K.~{Menou}, J.-M. {Hameury}, J.-P. {Lasota}, R.~{Narayan}, {\it \mnras\/} {\bf
  314}, 498 (2000).

\bibitem{owen2012}
J.~E. {Owen}, A.~P. {Jackson}, {\it \mnras\/} {\bf 425}, 2931 (2012).

\bibitem{kimura2016}
S.~S. {Kimura}, M.~{Kunitomo}, S.~Z. {Takahashi}, {\it \mnras\/} {\bf 461},
  2257 (2016).

\bibitem{knigge1999}
C.~{Knigge}, {\it \mnras\/} {\bf 309}, 409 (1999).

\bibitem{zhu2018}
Z.~{Zhu}, J.~M. {Stone}, {\it \apj\/} {\bf 857}, 34 (2018).

\bibitem{lijw2019}
J.~{Li}, X.~{Cao}, {\it \apj\/} {\bf 872}, 149 (2019).

\bibitem{zhu2022}
Y.~{Zhu}, D.-F. {Bu}, X.-H. {Yang}, F.~{Yuan}, W.-B. {Lin}, {\it \mnras\/} {\bf
  513}, 1141 (2022).

\bibitem{kong2021}
L.~D. {Kong}, {\it et~al.\/}, {\it \apjl\/} {\bf 906}, L2 (2021).

\bibitem{segura2022}
N.~{Castro Segura}, {\it et~al.\/}, {\it \nat\/} {\bf 603}, 52 (2022).

\bibitem{orosz1997}
J.~A. {Orosz}, R.~A. {Remillard}, C.~D. {Bailyn}, J.~E. {McClintock}, {\it
  \apjl\/} {\bf 478}, L83 (1997).

\bibitem{hameury1997}
J.~M. {Hameury}, J.~P. {Lasota}, J.~E. {McClintock}, R.~{Narayan}, {\it \apj\/}
  {\bf 489}, 234 (1997).

\bibitem{gallo2018}
E.~{Gallo}, N.~{Degenaar}, J.~{van den Eijnden}, {\it \mnras\/} {\bf 478}, L132
  (2018).

\bibitem{wangjy2021}
J.~{Wang}, {\it et~al.\/}, {\it \apjl\/} {\bf 910}, L3 (2021).

\bibitem{shaw2021}
A.~W. {Shaw}, {\it et~al.\/}, {\it \apj\/} {\bf 907}, 34 (2021).

\bibitem{kelly2007}
B.~C. {Kelly}, {\it \apj\/} {\bf 665}, 1489 (2007).

\bibitem{russell2006}
D.~M. {Russell}, {\it et~al.\/}, {\it \mnras\/} {\bf 371}, 1334 (2006).

\bibitem{paice2019}
J.~A. {Paice}, {\it et~al.\/}, {\it \mnras\/} {\bf 490}, L62 (2019).

\bibitem{blandford1979}
R.~D. {Blandford}, A.~{K{\"o}nigl}, {\it ApJ\/} {\bf 232}, 34 (1979).

\bibitem{kimura2021}
S.~S. {Kimura}, T.~{Sudoh}, K.~{Kashiyama}, N.~{Kawanaka}, {\it \apj\/} {\bf
  915}, 31 (2021).

\bibitem{aaz2015}
A.~A. {Zdziarski}, M.~{Sikora}, P.~{Pjanka}, A.~{Tchekhovskoy}, {\it \mnras\/}
  {\bf 451}, 927 (2015).

\end{thebibliography}


\section*{Acknowledgments}
We thank Wen-Fei Yu, Jian-Feng Wu, Fu-Guo Xie, Wei-Min Gu and Mou-Yuan Sun for the discussions. BY thanks Yong-Kang Zhang from Jinan University for the help in producing the animation. 
{\bf Funding:} 
B.Y. is supported by the National Program on Key Research and Development Project 2021YFA0718500; by NSFC grants 12273026 and U1931203; by Natural Science Foundation of Hubei Province of China 2022CFB167; by the Fundamental Research Funds for the Central Universities 2042022rc0002; by Xiaomi Foundation / Xiaomi Young Talents Program.
X.C. is supported by the NSFC grants 11833007, 12073023, 12233007, and 12147103; by the science research grants from the China Manned Space Project with No. CMS-CSST- 2021-A06; by the fundamental research fund for Chinese central universities (Zhejiang University).
Z.Y. is supported by the Natural Science Foundation of China grants U1838203 and U1938114; the Youth Innovation Promotion Association of CAS id 2020265; by the funds for key programs of Shanghai astronomical observatory.
S.-N.Z. is supported by the National Program on Key Research and Development Project Grant No. 2016YFA0400802; by International Partnership Program of Chinese Academy of Sciences (Grant No.113111KYSB20190020).
P.D. is supported by NSFC grants 12022301 and 11991051.
{\bf Author contributions:}
B.Y. initiated the project, performed the data analysis and model interpretation, and led the writing of the text. X.C. led the interpretation of the MAD and the writing of the relevant text. Z.Y. contributed to the data analysis, model discussion, and the writing of the text. J.-M.H. provided the DIM code and contributed to the model discussion and the writing of the text. B.C. contributed to the discussion of the DIM. Y.W. and T.X. contributed to the simulation of the DIM and the data analysis. M.S., S.-N.Z., P.D. and P.Z. contributed to the model discussion and the writing of the text. 
{\bf Competing interests:} The authors declare no competing interests.
{\bf Data and materials availability:} 
All \textit{Insight}-HXMT data used in this work (Proposal ID: P0114661) are publicly available and can be downloaded from the official website: http://archive.hxmt.cn/. The radio luminosities are taken from source data which are publicly available \cite{bright2020}, but are corrected for the source distance $d = 2.96$ kpc in this work. All the reduced data (including LE/ME/HE light curves and X-ray/Radio/$V$-band luminosities), the scripts which perform the ICCF analysis and plot the figures, and the numerical codes for computing the magnetic field strength (MAD) and the disk instability (DIM), are available at https://github.com/Bei-You-BH/MAD, which are archived at Zenodo \cite{mad2023}. The algorithm \texttt{linmix} we used for fitting the correlations is available at https://github.com/jmeyers314/linmix.

\clearpage

\begin{figure}
\centering
\centering\includegraphics[scale=0.4]{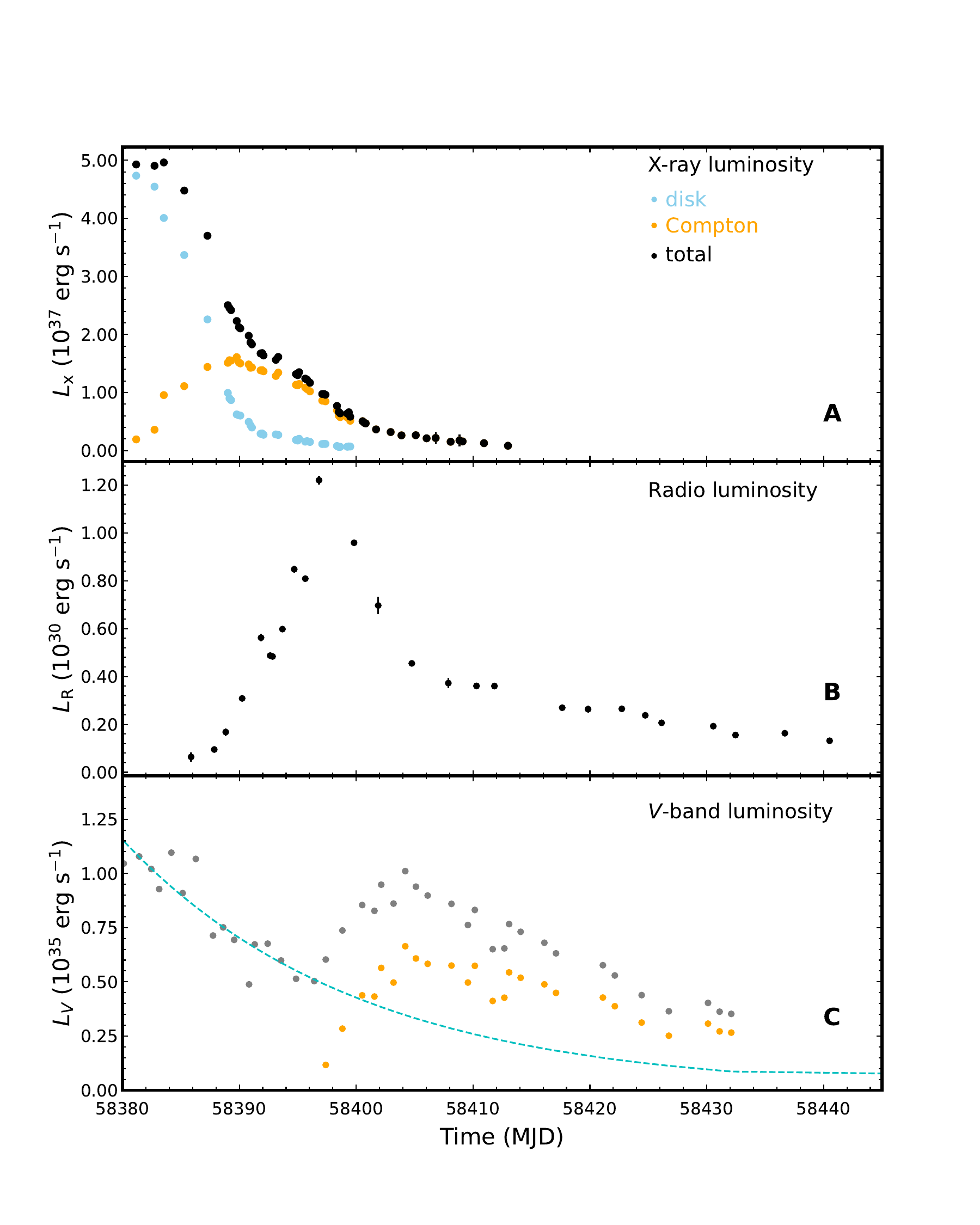}
\caption{
 Multi-wavelength light curves of the flare. A distance of $d = 2.96$ kpc \cite{atri2020} is adopted in estimating the observed luminosity. (A) The thermal luminosity from the thin disk, the Comptonization luminosity from the ADAF, and the total luminosity, in the 0.1-100 keV band, as determined by our decomposition of the \textit{Insight}-HXMT data \cite{methods}. (B) The radio emission at 15.5 GHz measured by AMI-LA. (C) The extinction-corrected $V$-band luminosities provided by AAVSO are represented by the gray points. The dashed cyan line is an exponential model fitted to the gray points between MJD 58380-58390. The orange points are the $V$-band luminosity between MJD 58395-58440 after subtracting the exponential model. The uncertainties, which are too small to be seen, are quoted at the 1-$\sigma$ level.   
 }
\label{multi_lc}
\end{figure}

\begin{figure}
\includegraphics[width=\linewidth]{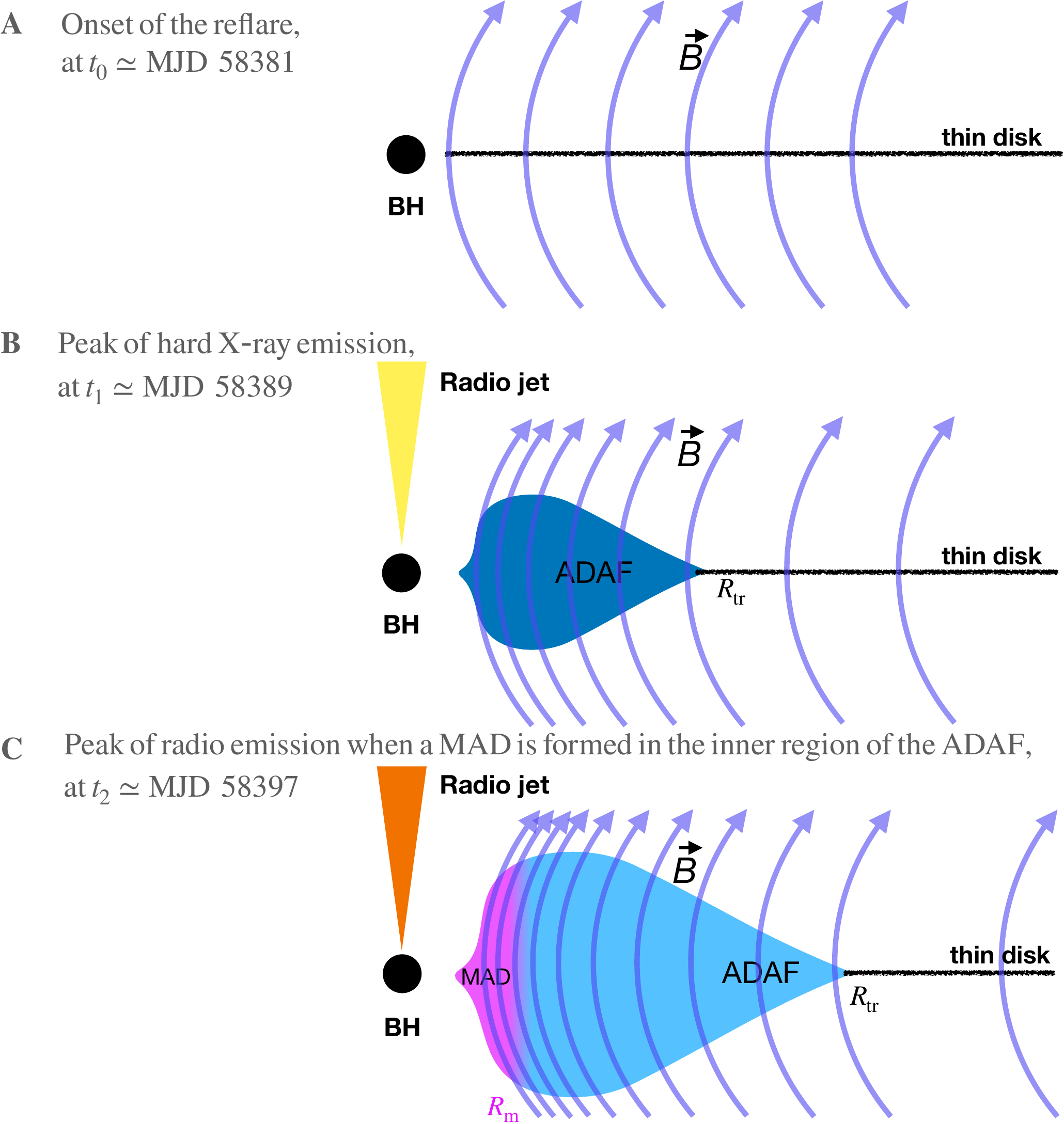} 
\caption{
Schematic diagram of our proposed interpretation. The outer thin disk is truncated at a radius $R_{\rm tr}$, within which is the ADAF. (A) The configuration at the onset of the flare, when the thin disk is assumed to extend to ISCO and the ADAF starts to expand its size with increasing $R_{\rm tr}$. (B) Configuration during the peak hard X-ray emission, when the outer thin disk brings both mass and magnetic field into the inner ADAF. (C) Configuration at the time of peak radio emission, when the sufficient magnetic field is archived through amplification by the ADAF. A MAD forms within the magnetospheric radius $R_{\rm m}$ where the magnetic pressure becomes equal to gravity force \cite{narayan2003}. Movie S1 shows an animated version of this figure.}
\label{schematic_MAD}
\end{figure}

\begin{figure}
\centering
\includegraphics[width=\textwidth]{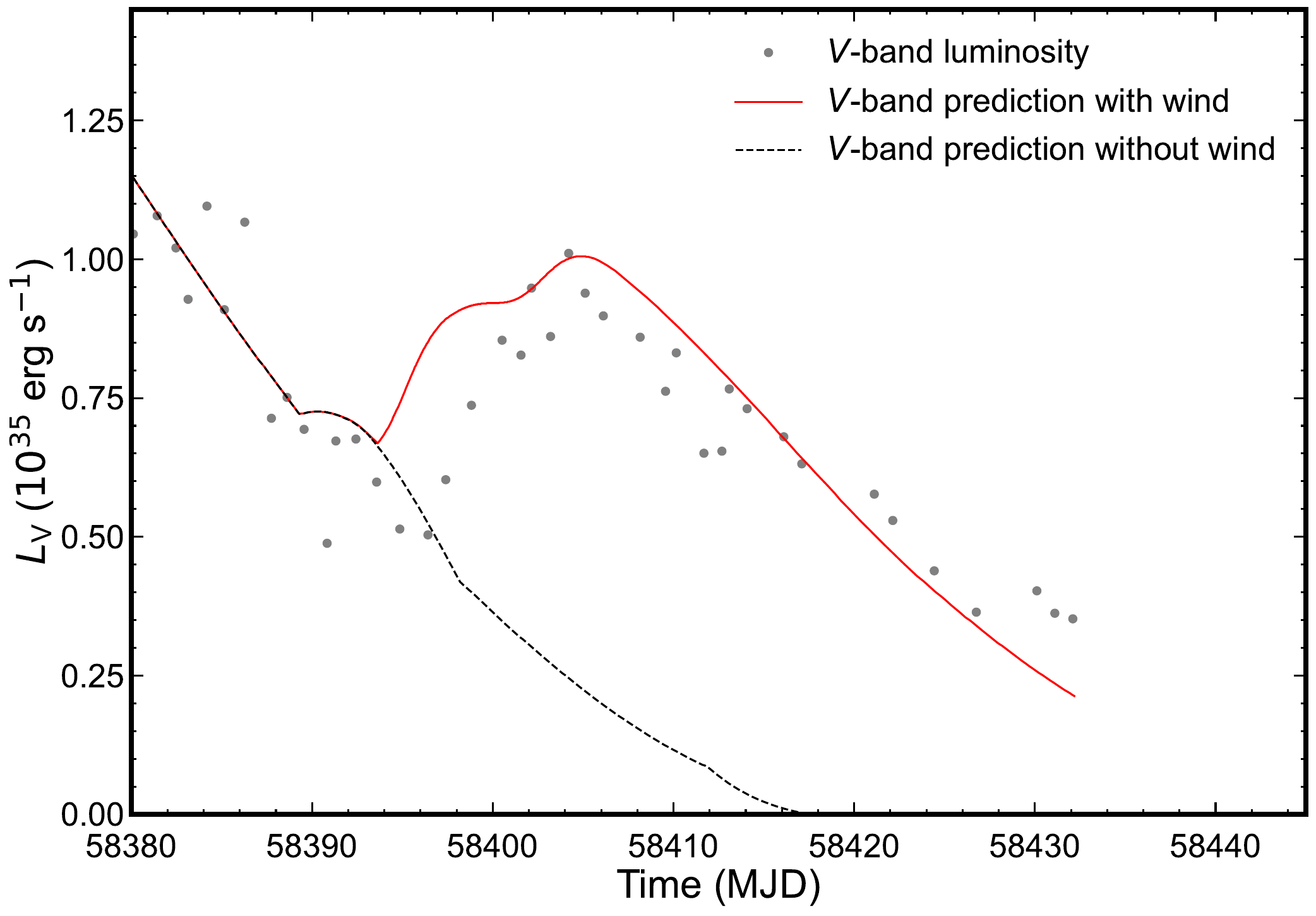} 
\caption{
 Observed $V$-band flare compared to our simulations. The $V$-band luminosity is from Fig. \ref{multi_lc}C (gray points). The red curve is the simulated $V$-band luminosity from the DIM simulation including the disk wind. The dashed orange line is the same equivalent simulation without the disk wind.
\label{flare}}
\end{figure}

\clearpage

\renewcommand{\thesection}{S\arabic{section}}
\renewcommand{\thetable}{S\arabic{table}}
\renewcommand\thefigure{S\arabic{figure}}
\renewcommand{\theequation}{S\arabic{equation}}
\pagenumbering{arabic} 
\setcounter{figure}{0}

 \begin{center}
{\Large Supplementary Materials for} \\
\vspace*{1cm}

\textbf{\large Observations of a black hole X-ray binary indicate formation of a magnetically arrested disk}

\vspace*{1cm}


{Bei You$^{\ast}$, Xinwu Cao$^{\ast}$, Zhen Yan$^{\ast}$, Jean-Marie Hameury, \\ 
Bozena Czerny, Yue Wu, Tianyu Xia, Marek Sikora, \\
Shuang-Nan Zhang, Pu Du, Piotr T. Zycki}

\baselineskip 24pt
\baselineskip 12pt

\vspace*{0.5cm}
{$^\ast$ Corresponding author. \\Email: youbei@whu.edu.cn, xwcao@zju.edu.cn, zyan@shao.ac.cn}

\end{center} 

\thispagestyle{empty}
 
 
\vspace*{2cm}

\noindent {\bf This PDF file includes:}\\

\noindent 
\hspace{1cm} Materials and Methods\\
\noindent\hspace*{1cm} Supplementary Text\\
\noindent\hspace*{1cm} Figures S1 - S8\\
\noindent\hspace*{1cm} Caption for Movies S1\\
\noindent\hspace*{1cm} Reference 41-97\\

\noindent {\bf Other Supplementary Materials for this manuscript :}\\

\noindent 

\noindent\hspace*{1cm} Movies S1

\clearpage


\section{Materials and Methods}
\subsection{Observations and data reduction} \label{sec:data}
The archival observation of MAXI J1820+070 by the \textit{Insight}-HXMT was performed from 2018 March 14 (MJD 58191) to 2018 October 21 (MJD 58412). The energy spectra from LE, ME and HE instruments are extracted following the same procedure as in the ref\cite{you2021}. The broad (2-150 keV) \textit{Insight}-HXMT energy spectra during the flare (MJD 58380--58412) are well fitted with a model \texttt{tbabs*(diskbb+nthcomp)} in \texttt{XSPEC} \cite{arnaud1996}, where \texttt{tbabs} is the absorption component with a fixed column density $N_{\rm H} = 0.15 \times 10^{22}$ cm$^{-2}$, \texttt{diskbb} is the multiple blackbody component from the thin disk, and \texttt{nthcomp} is the Comptonization component from the ADAF. As an example, the X-ray spectrum and the corresponding best-fitting model during the hard X-ray peak at MJD 58389 are plotted in Fig. \ref{spectrum_fit}. We exclude the energy range 21--24 keV, since the fluorescence lines around 21--24 keV due to the photoelectric effect of electrons in Silver K-shell are detected by the Si-PIN detectors of ME. Then, we use the \texttt{cflux} in \texttt{XSPEC} to estimate the unabsorbed disk and Comptonization luminosity, and the total luminosity in 0.1-100 keV during the flare. The disk luminosity continuously decreases with time, while the Comptonization luminosity peaks on MJD 58389 (Fig. \ref{multi_lc}A). The energy spectrum, i.e., $N(E) \propto E^{-\Gamma}$, becomes harder as the photon index $\Gamma$ decreases from about 2.4 to 1.6.

The reflection component is not included in the spectral fitting, since it is too weak to be detected by the  \textit{Insight}-HXMT during the flare. There are three observations during this flare by \textit{Nuclear Spectroscopic Telescope Array (NuSTAR)} with much better sensitivity \cite{buisson2019}. The reflection component in all the three \textit{NuSTAR} spectra is roughly 5-20 times weaker than the Comptonization component.

The radio flux (15.5 GHz) is directly obtained from the public work \cite{bright2020}, which is reduced from the observations by AMI-LA. We just use the new distance measurement of 2.96 kpc to recalculate the radio luminosity. The radio luminosity $L_\mathrm{R}$ during the flare is plotted in Fig. \ref{multi_lc}B.

The archival optical light curve is from the AAVSO. We select the $V$-band data with uncertainties $<$ 0.02 mag. We first converted the Johnson magnitudes to AB magnitudes \cite{frei1994}, and calculated the monochromatic flux according to $m_\mathrm{AB}=-2.5\ln f_\mathrm{V}-48.6$ \cite{oke1974}. We use $E(B-V)=0.23$ \cite{mikolajewska2022} to obtain the extinction correction flux. In order to study the flare on a daily basis, we rebined the $V$-band light curve with a bin size of 1 day, and the uncertainties are calculated by error propagation. The light curve in the $V$-band during the flare is plotted in Fig. \ref{multi_lc}C.


\subsection{Determination of time lags using the interpolated cross-correlation function}
The time lags between the radio/optical/X-ray light curves are measured using the ICCF which has previously been used in the time-series analysis of reverberation mapping of AGNs \cite{kaspi2000,bentz2009}. The correlation coefficient $r$ describes the strength of the correlation. The peak delays are obtained from the location with the maximum correlation coefficient $r_{\rm max}$, and the centroid delays are measured from the centroid of ICCF above a threshold ($r>0.8 r_{\rm max}$). The uncertainties of lags are determined by the flux randomization/random subset sampling (FR/RSS) method \cite{du2014}, which takes into account both the measurement uncertainties and the uncertainties resulting from the sampling cadence. In general, the FR/RSS method generates a subsample of the light curves by randomly selecting the points from the light curves and dithering the points according to their error bars. The centroid and peak lag distributions can be obtained from the ICCF of the subsamples. The confidence intervals of the lags can be determined using the centroid and peak lag distributions.

The $V$-band light curve contains the contribution of the underlying exponential decay from the main outburst as well as that of the flare. To subtract the contribution of the flux level from the underlying decay to this flare between MJD 58395 and 58440, we fit an exponential function between MJD 58380 and MJD 58390, then subtract it from the observed luminosity after MJD 58395. We use the subtracted light curve to estimate the lag between the X-ray and $V$-band flares.

We apply the ICCF to the multi-wavelength light curves in Fig. \ref{multi_lc}, and the results are plotted in Fig. \ref{58380_r_v_x}. We find that: i) the radio emission lags the X-ray emission by the peak time delay $\tau_{\rm RX}^{\rm p} = 8.06^{+0.88}_{-0.56}$ days and the centroid delay $\tau_{\rm RX}^{\rm c} = 8.32^{+1.31}_{-0.48}$ days; ii) the optical $V$-band emission lags the X-ray emission by the peak time delay $\tau_{\rm OX}^{\rm p} = 17.61^{+0.50}_{-0.44}$ days and the centroid delay $\tau_{\rm OX}^{\rm c} = 17.30^{+0.41}_{-0.56}$ days; iii) the optical $V$-band emission lags the radio emission by the peak time delay $\tau_{\rm OR}^{\rm p} = 8.61^{+1.06}_{-0.83}$ days and the centroid delay $\tau_{\rm OR}^{\rm c} = 8.80^{+0.64}_{-0.94}$ days. The peak and centroid time delays are consistent within uncertainties. We report the peak time delay of $\tau_{\rm RX}^{\rm p} = 8.06^{+0.88}_{-0.56}$ days, $\tau_{\rm OX}^{\rm p} = 17.61^{+0.50}_{-0.44}$ days and $\tau_{\rm OR}^{\rm p} = 8.61^{+1.06}_{-0.83}$ days in the main text.

\subsection{Formation of the MAD within the expanding ADAF}\label{sec:mad}

The LE X-ray light curve during the soft state between MJD $\sim$ 58330-58380 is used to estimate the mass accretion rate of the gas fed to the thin disk over time, i.e., $\dot{M}_{\rm d}(t)$.
We fitted the LE X-ray light curve during this soft state with an exponential function $\propto \exp[-(t-t_{\bullet})/\tau]$ \cite{king1998}, finding $t_{\bullet} = 58330.72$ and $\tau=57.61$ days. During the soft state, the thin disk approaches the ISCO of the BH, which indicates the mass accretion rate also decays exponentially with the same $t_{\bullet}$ and $\tau$\cite{king1998}. 


We assume that the disk luminosity $L_{\rm d}(t_{0}) = \eta \dot{M}_{\rm d}(t_{0})c^2$ at $t_{0}$ = MJD 58381, where the radiation efficiency $\eta=$0.1, a typical value adopted for a thin disk \cite{yuan2014}, and $c$ is the speed of light.
Then we estimate the Eddington-scaled accretion rate of the thin disk $\dot{m}_{\rm d} =\dot{M}_{\rm d}/\dot{M}_{\rm Edd}$ (Eddington accretion rate $\dot{M}_{\rm Edd} = L_{\rm Edd}/\eta c^2$, where Eddington luminosity $L_{\rm Edd} = 1.3\times 10^{38} M_{\rm BH}/M_{\odot} ~ \rm erg ~ s^{-1}$), which gives $\dot{m}_{\rm d}(t_{0}) =0.0474$. 
We assume $\dot{m}_{\rm d}(t)$ during the flare follows the same exponential decay as in the soft state \cite{lasota2001,tetarenko2018}, and then determine the mass accretion rates of the thin disk, $\dot{m}_{\rm d}(t_1)=0.041$ and $\dot{m}_{\rm d}(t_2)=0.036$.

With the mass accretion rate derived above and the observed disk luminosity provided in Section \ref{sec:data}, the truncation radius $R_{\rm tr}$ of the outer thin disk can be estimated.
The luminosity of a truncated disk is
\begin{equation}
L_{\rm d}(t)\propto {\frac {\dot{m}_{\rm d}(t)}{R_{\rm tr}(t)}}.
\label{lum_d}
\end{equation}
At $t=t_0$, we assume the disk extends to the ISCO $R_{\rm tr}(t_0)=R_{\rm ISCO}$. The truncated radius ${R_{\rm tr}(t)}$ for $t>t_0$ is estimated from the observed disk luminosity using
\begin{equation}
{\frac {R_{\rm tr}(t)}{R_{\rm ISCO}}}\simeq {\frac {\dot{m}_{\rm d}(t)L_{\rm d}(t_0)} {\dot{m}_{\rm d}(t_0)L_{\rm d}(t)}}.
\label{r_tr}
\end{equation}
We find that $R_{\rm tr}(t_1)/R_{\rm ISCO}\simeq 6.79$, and $R_{\rm tr}(t_2)/R_{\rm ISCO}\simeq 32.83$. The values of the truncation radius $R_{\rm tr}$ during the flare are plotted in Fig. \ref{R_tr}.

The Comptonization component predominantly originates from the inner hot ADAF, which we use to set constraints on the properties of the ADAF. Outflows are thought to be driven from the hot ADAF \cite{blandford1999,you2021}, and the mass accretion rate varies with radius; we model the Eddington-scaled accretion rate of the ADAF $\dot{m}_{\rm ADAF}$ as
\begin{equation}
\dot{m}_{\rm ADAF}(R)=\dot{m}_{\rm d}\left({\frac R{R_{\rm tr}}}\right)^{p_{\rm w}},
\label{mdot_r}
\end{equation}
where a typical value of $p_{\rm w} = 0.3$ is adopted to describe the radial dependence of the accretion rate of the ADAF due to the outflow \cite{xie2019}. Since the mass accretion rate $\dot{m}_{\rm d}(t_2)$ at the time when the radio emission peaked ($t=t_2$) is only slightly lower than $\dot{m}_{\rm d}(t_0)$, the radiation efficiency of the ADAF is roughly the same as that of a thin disk at $t=t_0$ \cite{yuan2014}. We therefore adopt a constant radiation efficiency. This implies that the Comptonization component emission is partly regulated by the outflows that reduce the mass accretion rate in the inner region of the ADAF, where most Compton scattered photons are emitted.

A fraction of gravitational energy released in the ADAF is carried away by the outflows. In the presence of a strong large-scale magnetic field threading the ADAF, we assume that such outflows are predominantly driven by the magnetic field. In this case, the magnetically driven outflows not only carry away the energy but also the angular momentum of the ADAF, which leads to an increase of the radial velocity of the accretion flow $v_{\rm R}$ by a factor $f_{\rm m}$,
\begin{equation}
v_{\rm R}=(1+f_{\rm m})v_{\rm R,\rm vis},
\label{v_r}
\end{equation}
where $v_{\rm R,vis}$ is the velocity of the pure viscously driven ADAF [\cite{cao2016}, their equation 16]. Only $\sim 1/(1+f_{\rm m})$ of the kinetic energy dissipated in the accretion flow is radiated out, while a fraction $\sim f_{\rm m}/(1+f_{\rm m})$ of the kinetic energy of the ADAF is tapped to accelerate the outflows \cite{cao2016}. Thus, the Comptonization luminosity normalized to the Eddington value, $\lambda_{\rm C}$ is approximated by
\begin{equation}
\lambda_{\rm C}(t)\equiv{\frac {L_{\rm C}(t)}{L_{\rm Edd}}}\sim {\frac {\dot{m}_{\rm d}(t)-\lambda_{\rm d}(t)}{1+f_{\rm m}}}
\left[  {\frac  {2.25R_{\rm ISCO}}{R_{\rm tr}(t)}}\right]^{p_{\rm w}},
\label{lum_comp}
\end{equation}
where $L_{\rm C}$ is the Comptonization luminosity, $\lambda_{\rm d}\equiv L_{\rm d}/L_{\rm Edd}$ is the observed Eddington ratio of the disk luminosity, and Equation \ref{mdot_r} is used. $\dot{m}_{\rm d} = \lambda_{\rm d}$ applies only in the condition of the disk not being truncated, i.e. $R_{\rm tr} = R_{\rm ISCO}$. In our scenario, the truncation radius increases with time, so $\dot{m}_{\rm d} > \lambda_{\rm d}$.
The radiation from the region with $R\sim 2.25R_{\rm ISCO}$ contributes to most of the accretion flow  luminosity \cite{shakura1973}. 
Using the derived $\dot{m}(t_2)=0.036$, $R_{\rm tr}(t_2)/R_{\rm ISCO}\simeq 32.83$, and the observed $L_{\rm C}(t_2)$, Equation \ref{lum_comp} gives $f_{\rm m}\simeq 0.92$. From Equation \ref{lum_d} and \ref{lum_comp}, the Comptonization luminosity $L_{\rm C} $ is approximately proportional to $ \dot{m}_{\rm d}(1-R_{\rm ISCO}/R_{\rm tr})R_{\rm tr}^{-p_{\rm w}}$, which depends on $\dot{m}_{\rm d}$ and the relevant $R_{\rm tr}$.
As $R_{\rm tr}$ increases with time, the Comptonization luminosity $L_{\rm C}$ increases because of $1-R_{\rm ISCO}/R_{\rm tr}$, while it decreases due to $R_{\rm tr}^{-p_{\rm w}}$ instead. Therefore, the dependence of the Comptonization luminosity on $\dot{m}_{\rm d}$ and $R_{\rm tr}$ leads to the observed hard X-ray flare.



The gas and magnetic field are fed into the inner ADAF from the outer thin disk \cite{bisnovatyi1974,vanballegooijen1989,lubow1994}. The delayed radio flare implies an increase in the jet power in this source after the X-ray flare.
For X-ray binaries, the magnetic field of the donor star could be the source of the external magnetic field, as has been proposed in the high-mass X-ray binary Cyg X-3 \cite{cao2020}. That system contains a Wolf-Rayet star, which could have a strong magnetic field with several hundred Gauss at its surface \cite{cao2020}. However, the magnetic field strength of the K5 companion star in MAXI~J1820$+$070 is likely a few Gauss \cite{donati2009}, which is too low for advection in the ADAF \cite{cao2019}. Therefore, the external magnetic field must be maintained by the outer thin disk.

We consider two mechanisms for generating a coherent magnetic field in the outer thin disk. The first one is the inverse-cascade of the dynamo process in a thin disk \cite{tout1996}. If the whole magnetic field loop generated in the outer disk is dragged by the ADAF, then no MAD would be formed near the BH, because no net flux is added to the ADAF.  
However, the size of magnetic field loops formed by this mechanism is about the disk radius \cite{tout1996}, so it is possible that the field line threading the inner ADAF connects to the outer thin disk as a whole loop. The field line threading the ADAF can then be advected to the BH, while the other end of the field line connects to the outer thin disk (or even diffuses out). A similar phenomenon has been observed in some numerical simulations \cite{beckwith2009}. The second potential mechanism is based on numerical simulations\cite{liska2020}. Those simulations found that the turbulence in a radially extended thick disk can generate large-scale poloidal magnetic flux in situ with a relatively strong initial toroidal magnetic field, which then accumulates to form a MAD state near the BH by accretion. The simulations were carried out for a thick disk; the magnetic field generation in thin disks is not as efficient as their thick counterparts \cite{liska2020}.

No quantitative relation between the disk properties and the large-scale magnetic field strength has been determined in numerical simulations for a thin disk. We therefore assume that a weak coherent magnetic field is generated through the inverse-cascade of the dynamo process in the outer disk. Thus, the magnetic field advection/diffusion in the ADAF can be calculated. The magnetic field strength $B_{\rm tr}$ in the region near $R \sim R_{\rm tr}$ of the outer thin disk is estimated as [\cite{cao2021}, their equation 27]
\begin{equation}
B_{\rm tr}\sim 2.48\times 10^{8} \alpha^{-1/20}m^{-11/20}\dot{m}_{\rm d}^{3/5} R_{\rm tr}^{-49/40}~{\rm Gauss}, \label{b_pd_gas}
\end{equation}
where $m = M_{\rm BH}/M_{\odot}$, and $\alpha$ is the viscosity parameter with a typical value $\sim 0.1$--0.4 \cite{king2007}.
The ratio of gas to magnetic pressure is
\begin{equation}
\beta={\frac {8\pi p_{\rm gas}}{B_{\rm tr}^2}}=2.533\times10^3\alpha^{-4/5}m^{-2/5}\dot{m}_{\rm d}^{1/5}R_{\rm tr}^{-1/10},
\label{beta}
\end{equation}
where $p_{\rm gas}$ is the gas pressure. Using the derived values of the mass accretion rate in the disk, we estimate $\beta\sim 2.6\times 10^3$ at $t_1$, and $\beta\sim 2.1\times 10^3$ at $t_2$ for $\alpha=0.1$. This implies that the field generated in the outer thin disk is weak.

In principle, we can calculate the magnetic field advection/diffusion based on a global solution of the ADAF with magnetically driven outflows \cite{lubow1994,cao2011}. 
The dynamics of the ADAF can be described by a self-similar solution \cite{narayan1994}, from which the radial velocity of a pure viscously driven ADAF $v_{R,\rm vis}\simeq-3\nu/2R$, where $\nu$ is the turbulent viscosity.   
The relative disk thickness $H/R$ of the ADAF is $\sim 1$ \cite{narayan1994}, which is adopted to calculate $v_{R,\rm vis}$ and then the radial velocity of the ADAF with magnetic outflows using Equation \ref{v_r}. With the derived radial velocity of the ADAF, the field advection and diffusion in the ADAF can be calculated when the value of the Prandtl number and the external coherent magnetic field strength are specified \cite{lubow1994,cao2011}. The magnetic Prandtl number $P_{\rm m}\equiv \nu/\eta_{\rm m}$ ($\eta_{\rm m}$ is the magnetic diffusivity) is found to be around unity, or $\sim 2-5$ with shearing box simulations \cite{parker1979,fromang2009,guan2009,lesur2009}.

In the inner region of the ADAF, the magnetic field needs to be enhanced until it dominates the dynamics, to produce the MAD \cite{igumenshchev2002,narayan2003}. We estimate the magnetic field strength $B_{\rm MAD}$ required for a MAD state \cite{narayan2003},
\begin{equation}
B_{\rm MAD}\sim 1.5\times 10^9(1-f_{\Omega})^{1/2}\epsilon^{-1/2}m^{-1/2}\dot{m}^{1/2}_{\rm ADAF}(R)R^{-5/4} ~~ {\rm Gauss}, \label{b_mad}
\end{equation}
where $\epsilon\equiv v_{\rm R}/v_{\rm K}\sim 10^{-2}$ to $0.1$ \cite{narayan2012,begelman2022,kawamura2022} and $v_{\rm K}$ is the Keplerian velocity; $f_\Omega=\Omega/\Omega_{\rm K}$ ($\Omega$ is the angular velocity and $\Omega_{\rm K}$ is the Keplerian one). In this case, we assume that the ADAF with magnetic outflows is arrested in the inner region (interior to a magnetospheric radius $R_{\rm m}$, where the magnetic pressure becomes equal to gravity force \cite{narayan2003}) of the ADAF, of which the radial velocity is described by
\begin{equation}
v_{R}=\left\{ \begin{array}{l}
        (1+f_{\rm m})v_{R,\rm vis}=-{\frac 3 2}(1+f_{\rm m})\alpha c_{\rm s}{\frac H R}, ~~~{\rm if}~~ R>R_{\rm m}; \\
         -\epsilon v_{\rm K},~~~ {\rm if}~~ R\le R_{\rm m},
\end{array} \right.
\label{v_r2}
        \end{equation}
where $c_{\rm s}$ is the sound speed. We calculate the field dragged inwards by the ADAF with the estimated value of $f_{\rm m}$, when the values of $H/R$, $\epsilon$ and $R_{\rm m}$ are specified\cite{cao2011}. The external coherent field strength is given by Equation \ref{b_pd_gas}, and we tune the value of $R_{\rm m}$ until the inner MAD satisfies the field strength required for a MAD (Equation \ref{b_mad}), if the magnetic field near the BH is sufficiently strong to form a MAD (see the blue solid line in Fig. \ref{b_r}). We note that, in most cases, the field strength is always lower than the MAD condition (Equation \ref{b_mad}) no matter what value of $R_{\rm m}$ is adopted, due to the weak external field generated in the outer thin disk or/and the low amplification of the field in the inner ADAF, which corresponds to an ADAF without entering a MAD state (see the red solid line in Fig. \ref{b_r}).

Fig. \ref{b_r} shows the magnetic field strength for $t_1$ and $t_2$ respectively. The Prandtl number $P_{\rm m}=2$ is assumed in our calculations. The rotational velocity of the gas in the MAD is found to be sub-Keplerian in the range of 0.5-1.0 \cite{begelman2022}, and we simply adopt $f_{\Omega}=0.5$ in the calculations. The parameters, $H/R=0.8$, and $\epsilon=10^{-2}$ are adopted in the calculations \cite{begelman2022,narayan2022}. Our calculations show that no MAD state is attained at $t_1$ (because the magnetic field strength of the ADAF is substantially lower than the MAD criterion). The strength of the magnetic field dragged inwards by the ADAF reaches {$\sim 7\times 10^7$}~Gauss near the BH at $t_2$ (the radio peak), and the inner ADAF ($R_{\rm m}=1.17R_{\rm ISCO}$) is magnetically arrested at $t_2$ (Fig. \ref{b_r}). 

The formation of a MAD in this flare is mainly caused by the substantially expanding ADAF (i.e., the receding outer thin disk) when the mass accretion rate declines by $\sim$24\%, although the detailed physical process triggering such an increase of $R_{\rm tr}$ by a factor of $\sim$ 30 is still unclear. 

The delay between the radio and hard X-ray indicates a continuous expansion of the ADAF after the hard X-ray peak. This allows for further amplification of the magnetic field in the ADAF. If the truncated radius changes marginally after the hard X-ray peak, the amplification of the field strength between $R_{\rm tr}$ and $R_{\rm ISCO}$ in the ADAF is almost unchanged, and therefore the field strength near the BH is roughly $\propto \dot{m}_{\rm d}^{3/5}$ (see Equation \ref{b_pd_gas}). The hard X-ray emission from the inner ADAF also depends on the mass accretion rate. In such a case, this produces nearly simultaneous variations in X-ray and radio bands for BHXRBs. 

\subsection{Jet power}
\label{sec:jet}

The jet power of MAXI J1820+070 during the rising hard state has been estimated as $\sim 10^{38}$ erg s$^{-1}$ by previous work \cite{tetarenko2021,zdziarski2022}. The radio luminosity of the jet at a given frequency correlates with the jet power as $L_{\nu}\propto P_{\rm jet}^{\frac{17}{12}-\frac{2}{3}\varsigma}$ \cite{falcke1996,coriat2009}, where the $\varsigma$ is the radio spectral index. We apply this correlation to estimate the jet power as $\sim 4\times10^{37}$ erg s$^{-1}$ at the radio peak of flare by using the observed radio luminosity at 15.5 GHz and assuming a flat radio spectrum ($\varsigma\sim$0). 

The powering of jets by the electromagnetic extraction of the rotating energy from the BH is called the Blandford-Znajek (BZ) mechanism \cite{blandford1977}.
The BZ power for a moderately spinning BH with $B_{\rm h}=7\times10^7$~Gauss is estimated to be $10^{37}$ erg s$^{-1}$ [\cite{livio1999}, their Equation 4] (see Fig. \ref{jet_power}), which is consistent with the jet power of MAXI J1820+070 at the radio peak (see above). 
This is also consistent with our interpretation that the expanding ADAF during the flare amplifies and transports the magnetic field from the outer disk, and eventually reaches the MAD level.

Our model focuses on the increase of magnetic field strength to form the MAD. The jet power we used is estimated from observation and consistent with BZ power from the magnetic field of the MAD. The model cannot determine how and when the jet is launched, so we cannot compare it to the onset of the radio emission.

\subsection{Disk instability model with the disk wind}\label{sec:dim}

The ICCF analysis between the X-ray/$V$-band flare indicates that, during the flare, the optical luminosity is due to viscous heating in the outer disk and not to reprocessing of X-rays. 
The hard X-ray flare illuminates the truncated disk, triggering the revival of the instability which propagates outwards in the outer disk and thus leads to the $V$-band flare.

We use a disk instability code\cite{hameury1998} to simulate the time-dependent evolution of the thin disk. The input parameters are: orbital period $P = 16.45$ hours \cite{torres2020}; primary and secondary masses $M_{\rm BH}= 8.5 M_{\odot}$, and $M_{\rm star}=0.6M_{\odot}$, respectively; mass transfer rate from the donor star $\dot{M}_{\rm tr}=5\times10^{16} \rm g \, s^{-1}$. 
The models with constant values of Shakura-Sunyaev viscosity $\alpha$ cannot reproduce the outburst light curves of dwarf novae \cite{smak1984}. Instead, two different viscosities for the  thin disk's thermal equilibrium at a given radius, i.e., $\alpha_{\rm h}$ for the hot regime, $\alpha_{\rm c}$ for the cold regime, were introduced in the DIM model \cite{hameury1998,lasota2001}, which is verified by the magnetohydrodynamic simulation \cite{hirose2014}. We assume $\alpha_{\rm h}=0.3$ and $\alpha_{\rm c}=0.02$ \cite{hameury1998,dubus2001}.
We assume that the thin disk is truncated at inner region due to evaporation at a rate of \cite{menou2000,dubus2001}
\begin{equation}
\dot{M}_{\rm eva}(R)=0.008\dot{M}_{\rm Edd}\left[ \left( \frac{R}{R_{\rm Sch}}\right)^{1/4} + 40\left( \frac{R}{800R_{\rm Sch}}\right)^2\right]^{-1}, 
\label{eq:evap}
\end{equation}
where $R_{\rm Sch}$ is the Schwarzschild radius. The truncation radius is determined by requiring that the evaporation rate is equal to the accretion rate, which varies with time. We set a lower limit of $5\times 10^8 \rm cm$ ($\sim 70R_{\rm Sch}$)\cite{menou2000} to avoid excessive computing time while keeping the timescales in the inner disk region much shorter than in the outer regions.
In principle, the disk instability is responsible for the full outburst and should account for the entire light curve starting from MJD 58192 to MJD 58440, including the flare. However, for simplicity, we do not generate the entire light curve, but instead use the DIM simulation to explore the disk evolution during the flare when a cooling front (appearing as a depression in the surface density profile) associated with the main outburst (before MJD 58380) has swept inwards over the thin disk. 

In the DIM, the disk evolution is governed by the vertically integrated radial equations for mass, angular momentum, and energy conservation \cite{hameury1998}. For the energy conservation, the surface heating and cooling rate are determined using a pre-calculated grid of vertical structures in the ($\Omega$, $\Sigma$, $T_{\rm c}$ and $T_{\rm irr}$) space \cite{hameury1998,dubus2001}. 
$\Omega$, $\Sigma$, $T_{\rm c}$ and $T_{\rm irr}$ are the angular velocity, local surface density at radius $R$, midplane temperature, and irradiation temperature respectively. $T_{\rm irr}$ is determined from the irradiation luminosity $L_{\rm irr}$ as
$\sigma T_{\rm irr}^{4}= L_{\rm irr}/4 \pi R^{2}$, where $\sigma$ is the Stefan-Boltzmann constant. In this approach, irradiation increases the surface disk temperature $T_{\rm s}$, according to $\sigma T_{\rm s}^4 = Q^+ + \sigma T_{\rm irr}^4$ where $Q^+$ is the viscous heating flux; this assumes that irradiated X-rays do not penetrate deep into the disk. We do not take into account other effects such as the evaporation of the disk or the formation of outflows \cite{owen2012,kimura2016}. Instead, they are approximated separately either via an explicit evaporation term in the inner disk (Equation \ref{eq:evap}), or via an explicit disk wind term (see below). Although not fully consistent, this simplified approach can model the outburst of BHXRB \cite{dubus2019}.

The irradiation luminosity is assumed to be related to the mass accretion rate at the inner disk radius $\dot{M}_{\rm in}$, as $L_{\rm irr} = \zeta_{\rm d}\dot{M}_{\rm in}c^2$, where $\zeta_{\rm d}$ includes both the radiative efficiency and irradiation efficiency.
However, during the flare, the observed hard X-ray emission originating from the ADAF is the major contributor to X-rays illuminating the outer thin disk, so that $L_{\rm irr} = \zeta_{c}L_{\rm X}$. This additional illumination heats the disk region behind the cooling front to be hot enough so that hydrogen begins to be reionized (referred to as a 'hot state'). Thus, the thin disk switches back to the hot state with the revival of the instability.
In this case, the resulting spike in the density profile and the steep gradients of the temperature profile (heating front) propagates outwards, which then leads to a lagged optical ($V$-band) light curve \cite{dubus2001}. 
Therefore, in our simulation, the irradiation luminosity is $L_{\rm irr} = \rm max(\zeta_{c}L_{\rm X}, \zeta_{d}\dot{M}_{\rm in}c^2 )$. The time-dependent $L_{\rm X}$ is estimated by fitting the observed Comptonization X-ray luminosity (Fig. \ref{multi_lc}B) with a Gaussian function.
The irradiation efficiency represents the fraction of the X-ray luminosity irradiating the outer thin disk, which depends on the geometry of the inner ADAF and of the outer thin disk, the X-ray albedo, and the X-ray spectrum. The irradiation efficiency 
has to be tuned to trigger the instability at a certain radius of the thin disk. For simplicity, in our DIM simulation, we assume a constant efficiency, $\zeta_{d} = 1\times 10^{-4}$ and $\zeta_{c} = 7.5\times10^{-5}$ following previous work\cite{dubus2001}.

We introduce the disk wind into the disk instability model following a previously described method \cite{dubus2019}. The continuity equation and specific angular momentum equation are:
\begin{equation}
\frac{\partial \Sigma}{\partial t}-\frac{1}{2 \pi R} \frac{\partial \dot{M}}{\partial R}=-\dot{\Sigma}_{\mathrm{w}},
\label{continuity}
\end{equation}

\begin{equation}
\Sigma \frac{\partial R^{2} \Omega}{\partial t}-\frac{\dot{M}}{2 \pi R} \frac{\partial R^{2} \Omega}{\partial R}=\frac{1}{R} \frac{\partial}{\partial R}\left(R^{3} \nu \Sigma \frac{\partial \Omega}{\partial R}\right)-j \dot{\Sigma}_{\mathrm{w}},
\label{angular_momentum}
\end{equation}
where $\dot{M}$ is the mass accretion rate at radius $R$, and $\Omega$ is assumed to be Keplerian in our simulations. $\nu$ is the kinematic viscosity coefficient, which is determined by the Shakura-Sunyaev prescription $\nu = \alpha c_{\rm s}H$ where $c_{\rm s}$ is the sound speed, $H$ is the disk scale height and $\alpha$ is a coefficient equal to $\alpha_{\rm c}$ on the cold branch when hydrogen is neutral and to $\alpha_{\rm h}$ on the hot branch when hydrogen is ionized.
$\dot{\Sigma}_{\mathrm{w}}$ is the loss rate of the local surface density due to the disk wind, which is assumed to be related to the local surface density $\Sigma$ via a factor $f$ as $\dot{\Sigma}_{\mathrm{w}}=f\Sigma$, where $f$ is in units of $\rm s^{-1}$. $j$ is the excess specific angular momentum carried away by the disk wind, which is assumed to be proportional to the Keplerian specific angular momentum, i.e. $j=g\sqrt{GM_{\rm BH}R}$, where $g$ is the factor of setting the excess specific angular momentum carried away by the disk wind. We assume the factor also depends on the radius as $g = g^{'}R_{9}$ [$R_{9}\equiv R/(10^{9} ~\rm cm)$], where $g^{'}$ is a scaling factor.

In the case of the thermal disk winds, no excess angular momentum is carried by the disk winds so $g=0$\cite{knigge1999}, while in the case of magnetic disk winds, $g>1$ is expected in the global ideal magnetohydrodynamics (MHD) simulations of thin disks \cite{zhu2018}. In our simulations, the wind mass loss rate 
$\dot{\Sigma}_{\mathrm{w}} = \partial \Sigma / \partial t $ is $\sim 10^{-4} ~\rm g \mathrm{~cm}^{-2} \mathrm{~s}^{-1}$ 
and $\Sigma \sim 10^2 \mathrm{~g} \mathrm{~cm}^{-2}$. Setting $\dot{\Sigma}_{\mathrm{w}}=f\Sigma$, the factor $f$ is $\sim 10^{-6} ~~\rm s^{-1}$.

During the flare, no spectroscopic data are available to put constraints on the disk wind loss and its evolution. Therefore, we parameterize the disk wind in terms of the mass loss (i.e. the factor $f$) and the removal of the angular momentum (i.e. the factor $g$). We assume that the mass loss due to the disk wind is time- and radius-dependent as
\begin{equation}
f(t,R)=f^{'}(t)R_{9}\!^{-1}\frac {\alpha(R)-\alpha_{\rm c}} {\alpha_{\rm h}-\alpha_{\rm c}} ~ \rm s^{-1},
\end{equation}
where the term [$\alpha(\rm R)-\alpha_{\rm c}]/[\alpha_{\rm h}-\alpha_{\rm c}]$ ensures that the disk wind is driven in the hot branch only. The $R_{9}$ dependence means that the disk wind characteristic timescale $1/f$ increases with radius. 
We assume that the disk wind starts at $t_{\rm start}$ with $f^{'} = 0$, and mass loss rapidly increases until $t_{\rm end}$, following
\begin{equation}
f^{'}(t)= \left\{\begin{aligned}
0 &,\quad\text{if $t\leq t_{\rm start}$}.\\
\frac{f^{'}_{\rm max}}{t_{\rm end}-t_{\rm start}}\left(t-t_{\rm start}\right) &,\quad\text{if $t_{\rm start}\leq t<t_{\rm end}$}.\\
f^{'}_{\rm max}&,\quad\text{if $t \geq t_{\rm end}$.}
\end{aligned}
\right.
\end{equation}
The linear dependence of both $f$ and $g$ on the radius is not a unique solution to the disk wind for the optical flare. According to the definitions above, the loss rate of the angular momentum per unit area via the disk wind is determined by the term $fg\Sigma R^{1/2} \propto fgR^{-1/4}$ where $\Sigma \propto R^{-3/4}$ for a Shakura \& Sunyaev disk. If the term $fg$ is independent of radius, the loss rate of the angular momentum will be low at a large radius. We also try other options with different scaling forms, e.g. $f \propto 1/R_{9}^2$, which can also reproduce the observed light curve. We adopt the linear radial dependence of $g$ in the following simulations.

Using the above parameterization of the disk wind, the DIM simulations show that the revival of the instability is triggered at $R=5\times 10^{10}$ cm ($\sim 3.9\times 10^4 R_{\rm g}$, where $R_{\rm g} = GM_{\rm BH}/c^2$ is the gravitational radius), the heating front reached a maximum distance of $R=7\times 10^{10}$ cm ($\sim 5.6\times 10^4 R_{\rm g}$),
and the predicted $V$-band light curve roughly matches the observed one (Fig. \ref{flare}), when $t_{\rm start}$ = MJD 58391, $f^{'}_{\rm max} = 2.7\times 10^{-7} ~ \rm s^{-1}$ (i.e., 1/42.9 $\rm day^{-1}$) at $t_{\rm end}$ = MJD 58395 and $g^{'}=3.167$. The simulated $V$-band light curve lags the X-ray light curve by more than 15 days, in agreement with the observations.

The disk wind can explain the delay of the optical peak. The peak position of the optical emission in the light curve depends on how far away the heating front can reach.
For a normal outburst, 
in which the heating front reaches the outer edge of the thin disk, the initial rise from quiescence and outburst peak are unaffected by the existence of disk winds [\cite{tetarenko2018}, their Fig. 3].
By contrast, in our DIM simulations, the propagation of the instability during the flare starts from the soft state, when the cooling front is already propagating inwards over the disk. Due to the illumination by the hard X-ray flare of the truncated thin disk, the disk region behind the cooling front is heated until it is hot enough for the hydrogen to be reionized. Thus, the thin disk switches back to the hot state with a revival of the instability propagating outwards in the outer thin disk.

We postulate that the disk wind is launched in regions where the thin disk is in the hot state during the flare. The disk wind carries angular momentum ($g$ is non-zero) which is equivalent to an increase in the viscosity. Because the disk wind is launched only in the hot regions of the thin disk, the increase is in $\alpha_{\rm h}$, not $\alpha_{\rm c}$. An increased $\alpha_{\rm h}$ leads to a decrease in the minimum density required for the hot state at a given radius, so the heating front propagates further out, i.e. the optical luminosity increases for a longer period.


To illustrate the role of the disk wind in moving the heating front further out, we carry out a simulation without disk wind, i.e., setting $f=0$. A flare still appears, starting at around MJD 58390 (Fig. \ref{flare}), but the heating front is not strong enough to propagate at very large distances in the outer disk, and the resulting $V$-band flare fades away in a few days. This does not match the observations.

The wind launched from the thin disk has been studied numerically in previous work\cite{lijw2019,zhu2022}. Observations have shown that there are disk winds in the soft state \cite{you2016} and in the hard state \cite{kong2021,segura2022} for accreting systems.
The thin disk wind was present throughout the outburst of MAXI J1820+070 in 2018, as indicated by infrared spectroscopy \cite{sanchez2020},
although only two spectra are available during the flare (at MJD 58390 and 58391).
We included the disk wind during the flare only, ignoring the role of a pre-existing disk wind in the rising hard state and soft state. A pre-existing disk wind can be approximated in the model using a viscosity parameter larger than the pure magnetorotational instability (MRI) value \cite{tetarenko2021} - which is not well determined yet.
The mass loss rate via the disk wind and its time evolution cannot be determined from the available observational data \cite{sanchez2020}. Therefore, we cannot fully include the disk wind in the DIM simulation and do not explicitly include the disk wind prior to the flare. We do require, however, that the wind mass loss rate increases during the flare.

In another BHXRB GRO J1655-40, an optical rise occurred a few days before the X-rays \cite{orosz1997}, due to the truncation of the thin disk. Modeling indicated that the instability developed in a truncated disk, and the inner disk builds up on viscous timescale to emit X-rays, leading to the lagged X-ray emission in GRO J1655-40 \cite{hameury1997}. 
By contrast, in our interpretation of the onset of the soft-to-hard transition of MAXI J1820+070, the thin disk has been built up and extended to the ISCO. As the truncation radius of the outer thin disk recedes with decreasing mass accretion rate, the ADAF gradually builds up, leading to the hard X-ray flare occurring first. This X-ray flare illuminates the truncated thin disk, triggering the revival of the instability which propagates outwards in the outer disk, producing the lagged optical flare.

\subsection{Correlations between radio/$V$-band and X-ray }

\subsubsection{Correlation between radio and X-ray}\label{corre_x_r}


There is a universal radio-X-ray (R-X) correlation $L_\mathrm{R}\propto L_\mathrm{X}^{\Gamma}$ with $\Gamma \sim 0.6$ in the BHXRBs \cite{gallo2018}, which indicates strong accretion-jet coupling. Previous R-X correlation analysis usually used the 1--10 keV (or 3--9 keV) X-ray luminosity. In this work, we investigate the R-X correlation by using the Comptonization luminosity $L_\mathrm{C}$ instead.

$L_\mathrm{R}$ and $L_\mathrm{C}$ show no time delay during the rising hard state (MJD 58191-58283), according to the ICCF analysis. Note that, after the peak of the rising hard state (MJD 58200), the X-ray corona is contracting over time while the truncation radius of the thin disk is close to ISCO and remains unchanged \cite{kara2019}. The same conclusion is also drawn from the follow-up detailed spectral fits of \textit{NuSTAR} \cite{buisson2019}, \textit{Neutron star Interior Composition Explorer (NICER)} \cite{wangjy2021} and \textit{Insight}-HXMT \cite{you2021} observations. 
As discussed in Section \ref{sec:mad}, the increase of the truncation radius during the flare (after MJD 58380) leads to the observed 8-day radio lag relative to the hard X-ray emission. Therefore, the constant truncation radius of the thin disk during the rising hard state may be responsible for the observed zero delay between the X-ray emission and the radio emission.

We then find that the simultaneous $L_\mathrm{R}$ and $L_\mathrm{C}$ (time difference less than 1 day) follow a correlation $L_\mathrm{R}\propto L_\mathrm{C}^{0.53\pm0.06}$ during the rising hard state (blue dashed line in Fig. \ref{xr}), which is consistent with that in ref.\cite{shaw2021}. We use the \texttt{linmix} algorithm \cite{kelly2007} implemented in Python to determine the above correlation by fitting the data. During the flare, the data during the soft-to-hard state transition apparently deviate from the above correlation, and then roughly follow it after the transition to the decaying hard state (see Fig. \ref{xr}). 

 

\subsubsection{Correlation between X-ray and $V$-band}\label{corre_x_v}
There is also a correlation between X-ray and optical luminosities in the hard states of BHXRBs, which has been used to infer the origin of the optical emission \cite{russell2006}. We first investigate the  $L_\mathrm{V}$  and $L_\mathrm{C}$ during the rising hard state. There is no time delay detected between them, according to the ICCF analysis. The Spearman coefficient of the simultaneous $L_\mathrm{V}$ and $L_\mathrm{C}$ (time difference less than 1 day) is $0.92\pm0.01$ at a significance of 10.98 $\sigma$. We then fit the correlation with a linear function on the logarithmic scale and find $L_\mathrm{V} \propto L_\mathrm{C}^{1.00\pm0.09}$. This slope is steeper than the universal $L_\mathrm{V}$-$L_\mathrm{X}$ correlation in the hard state of BHXRBs ($\sim$0.6) \cite{russell2006}.

We then investigate the correlation between $L_\mathrm{V}$  and $L_\mathrm{C}$ during the flare. The Spearman coefficient is $-0.33\pm0.02$ at a significance 1.7 $\sigma$, which indicates there is no correlation between $L_\mathrm{V}$ and $L_\mathrm{C}$. Apparently, $L_\mathrm{V}$ and $L_\mathrm{C}$ during the flare (both in the soft-to-hard state transition and decaying hard state) do not follow the same positive correlation as that during the rising hard state (see Fig. \ref{xv}). 


\section{Supplementary text}

\subsection{Radio and optical delay due to jet traveling}
We also notice that the optical fast variability lags behind the X-ray by roughly 165 ms in MAXI J1820+070 [\cite{paice2019}, see also the time delay of radio variability in ref.\cite{tetarenko2021}]. Moreover, the delay increases with wavelength as 3.25 $\mu {\rm s} ~ \AA^{-1}$ \cite{paice2019}. Since the shorter wavelength emission is expected from the inner part of the jet \cite{blandford1979}, this time delay of fast variability and the wavelength dependence can be well explained by the propagation of perturbations in the jet and gives the jet bulk Lorentz factor as $6.8$ \cite{tetarenko2021}. If we linearly extrapolated the wavelength to 15.5 GHz, the time lag is $\sim 10$ minutes, which is much smaller than the derived radio lag of $\sim$ 8 days in our work. Therefore, our extremely long optical/radio delays cannot be associated with the jet traveling time. 

\subsection{MAD of BHXRB in quiescent}
According to Equation \ref{b_mad}, the magnetic field strength for a MAD depends on the mass accretion rate $\dot{m}$ and truncation radius $R_{\rm tr}$, which were both estimated from the observed disk component in our model (Equation \ref{r_tr} and Equation \ref{mdot_r}). 
The time variations of the magnetic field strength at the inner edge of the ADAF are shown in Fig. \ref{b_t}. We find that the magnetic field strength decreases after $t=t_2$, which is caused by the decreasing MAD field strength due to the decrease of the mass accretion rate over time (see Equation \ref{b_mad}). 
Nevertheless, based on the physical quantities $\dot{m}_{\rm d}$ and $R_{\rm tr}$ derived from the X-ray light curves (the disk and Comptonization component emissions), our model calculations show that the magnetic field strength in the ADAF peaks at a time corresponding to the maximal radio emission, and then declines. The disk emission is too weak to be detected after MJD$~$58400 when radio emission was still decreasing. This prevents us from determining $\dot{m}$, $R_{\rm tr}$, and therefore the magnetic field strength (Fig. \ref{b_t}), using the methods above \cite{methods}. But, it was also shown in previous work that the MAD would exist in the quiescent state \cite{kimura2021}.

\subsection{Standard and normal evolution (SANE) accretion flow}

Previous work has argued that the broadband X-ray spectrum of a MAD is quite similar to that of a SANE accretion flow \cite{xie2019}, which proposes that these two different types of accretion flows are indistinguishable using only X-ray spectroscopy. 
However, for the MAD with very low accretion rate $\dot{m} < 10^{-5}$, magnetic reconnections heat up the thermal/nonthermal electron, and the Thomson optical depth of MADs is too low for X-rays to be emitted by Comptonization. Thus, it was proposed that a MAD emits nonthermal synchrotron emission which dominates over Comptonization, and produces a hard and power-law spectrum in X- and gamma-ray bands \cite{kimura2021}. In contrast, the SANE-mode accretion flow can emit hard X-ray emissions via Comptonization. Therefore, the SANE-mode accretion flow leads to soft spectra in the hard X-ray range, while MADs produce power-law hard X-ray spectra owing to the nonthermal electrons. In addition, the radio luminosity of MADs may be different from that of SANE-mode accretion flows because of the difference in magnetic fields \cite{aaz2015}. 

\clearpage

\renewcommand{\thefigure}{{S1}}
\begin{figure}
\includegraphics[width=\linewidth]{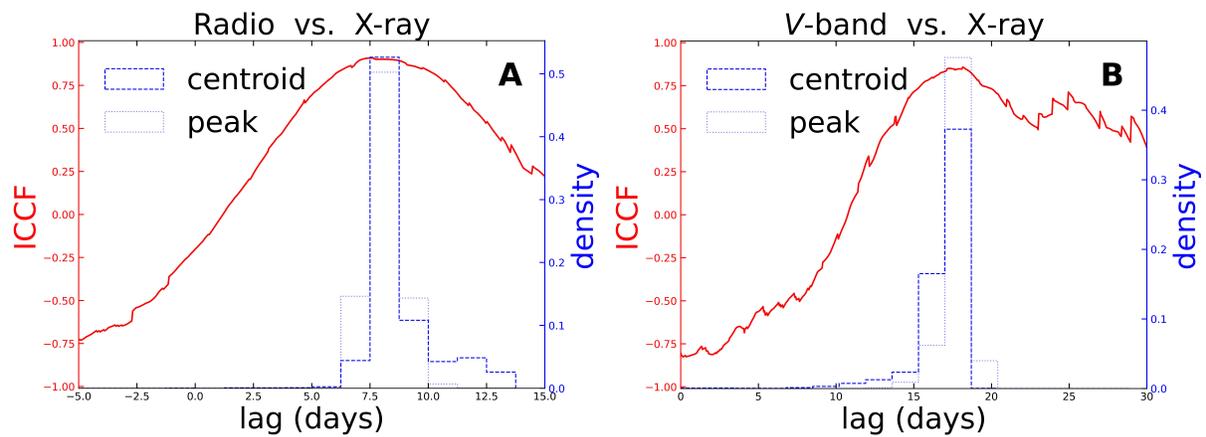}
\caption{
Cross-correlation analysis between the (A) radio and the X-ray luminosities and (B) $V$-band and the X-ray luminosities, both after MJD = 58380. In both panels, the  red line shows the cross-correlation function (left axis) for each time lag tested. The dashed and dotted blue histograms are the cross-correlation centroid lag distribution and the peak lag distribution respectively (right axis). See text for details. 
}\label{58380_r_v_x}
\end{figure}  

\renewcommand{\thefigure}{{S2}}
\begin{figure}
\includegraphics[width=\linewidth]{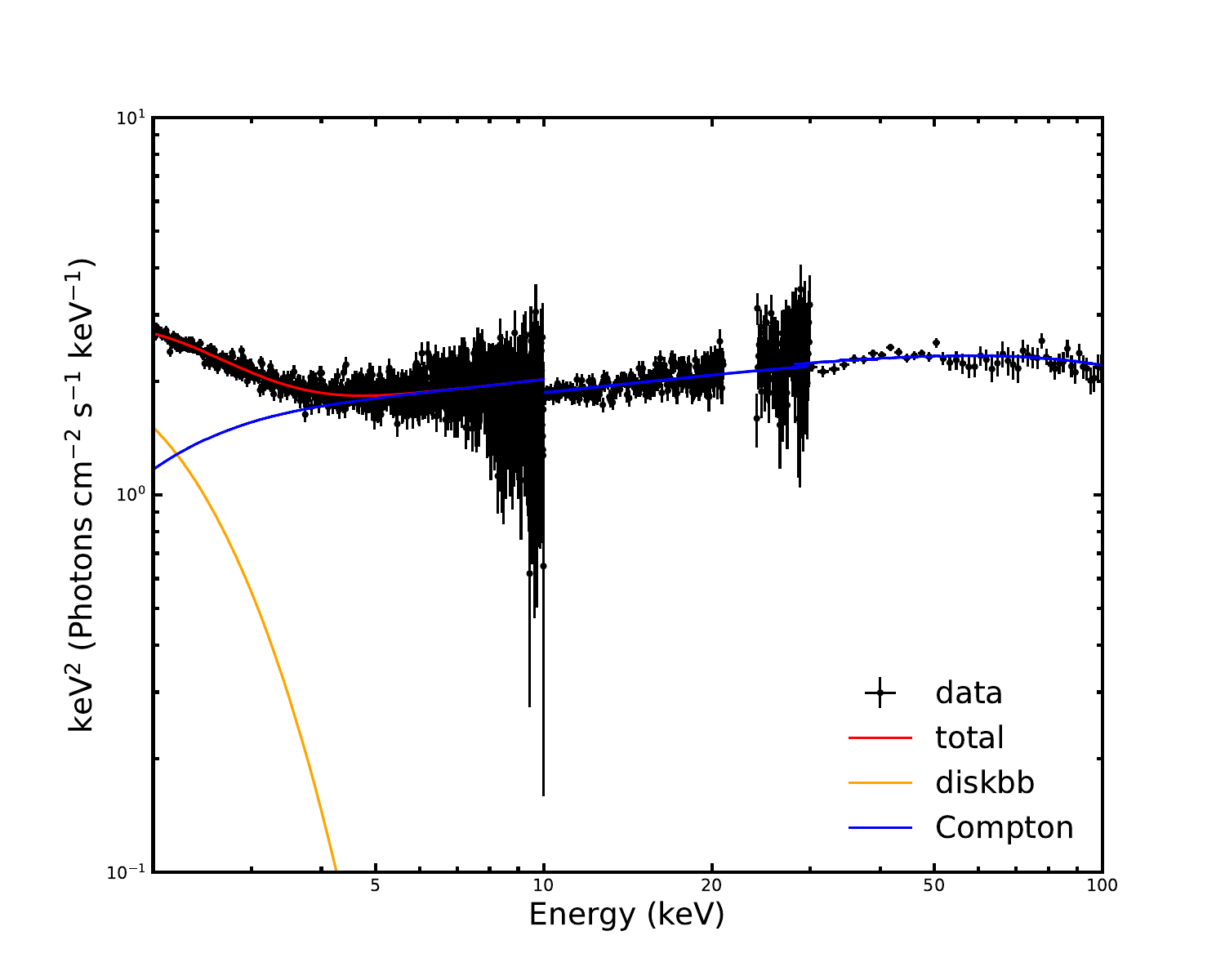} 
\caption{
Example unfolded energy spectrum of MAXI J1820+070 (black data points) fitted with the model \texttt{tbabs*(diskbb+nthcomp)}, during the hard X-ray peak at MJD = 58389. \texttt{tbabs} is the absorption component with a fixed column density $N_{\rm H} = 0.15 \times 10^{22}$ cm$^{-2}$, \texttt{diskbb} is the multiple blackbody component from the thin disk, and \texttt{nthcomp} is the Comptonization component from the ADAF. The total model is plotted in red, which is the total of the \texttt{diskbb} in orange and the \texttt{nthcomp} in blue. We exclude the spectrum in the range of 21-24 keV due to contaminating atomic lines. The error bars of the spectrum correspond to a 68\% confidence interval.
}\label{spectrum_fit}
\end{figure}

\renewcommand{\thefigure}{{S3}}
\begin{figure}
	\centering
	\includegraphics[width=\linewidth]{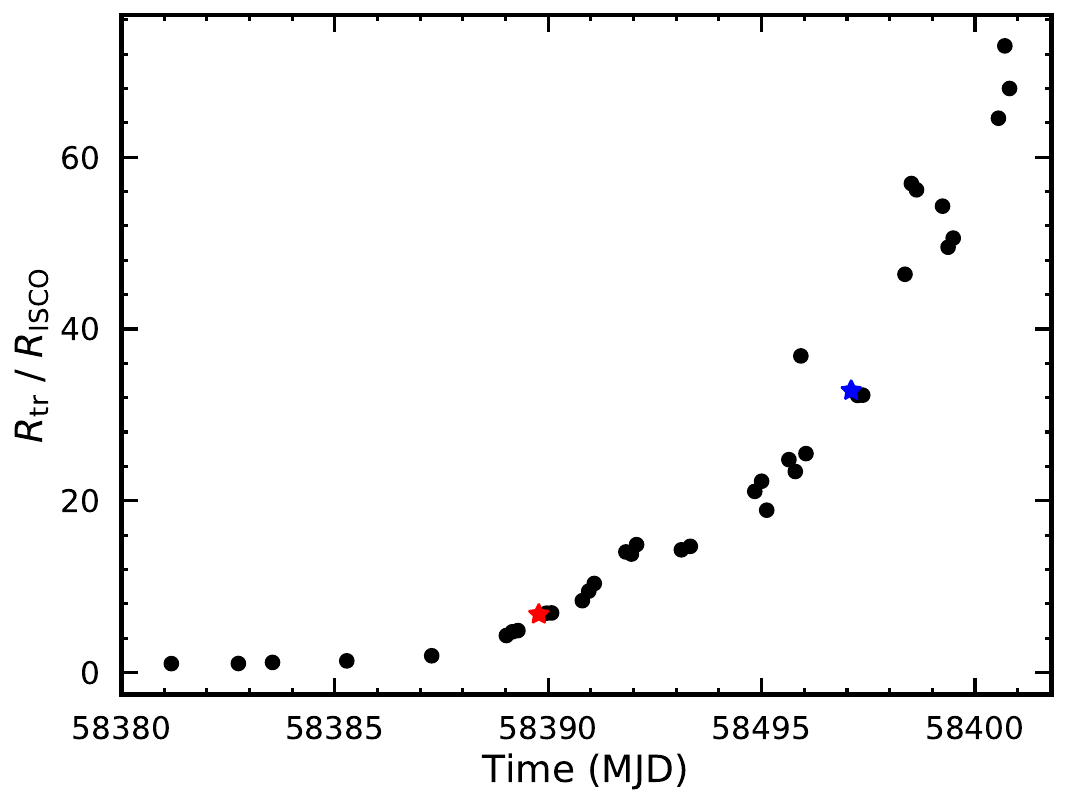}
	\caption{The truncation radii derived from the disk component of the X-ray emission with Equation \ref{r_tr}. The red and blue stars represent the truncation radii at $t=t_1$ and $t=t_2$ \cite{methods}, respectively. The same method is also applied to other epochs of \textit{Insight}-HXMT observation to estimate the truncation radii during the flare (black points).}
	\label{R_tr}
\end{figure}

\renewcommand{\thefigure}{{S4}}
\begin{figure}
	\centering
	\includegraphics[width=\linewidth]{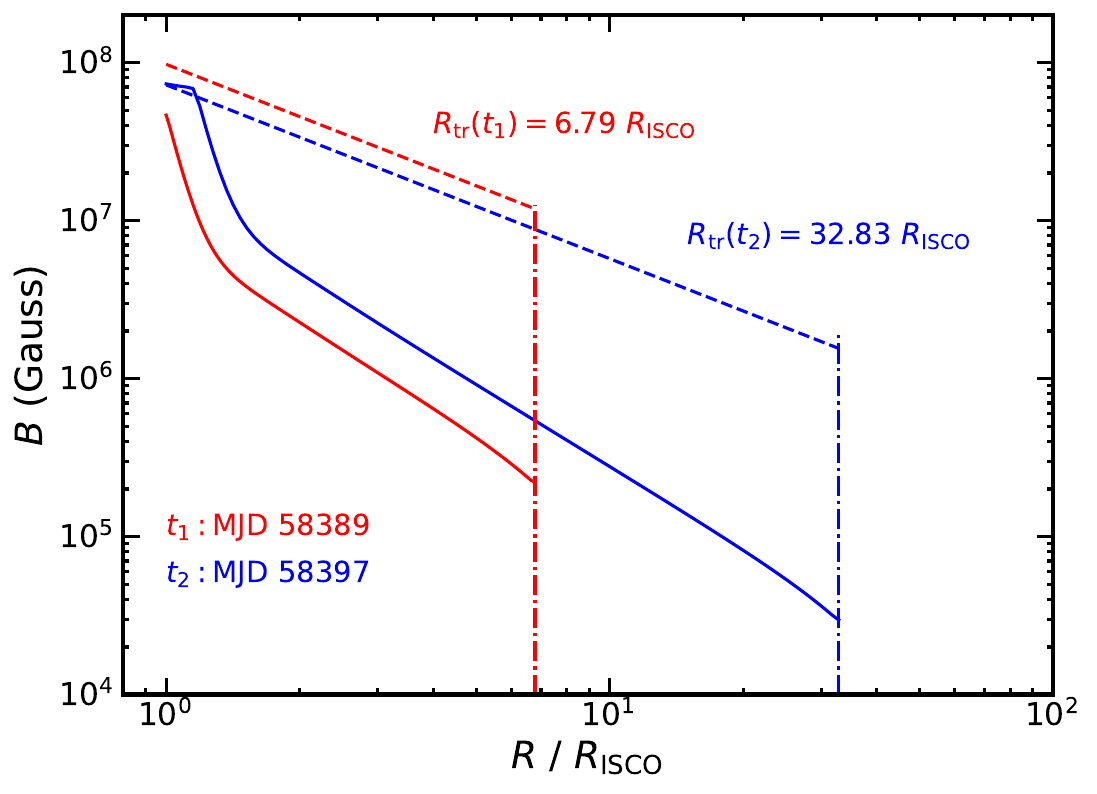}
	\caption{Strength of the magnetic field advected by the ADAF in the simulations as a function of radius (solid lines). The dashed lines are the required strength to satisfy the MAD criterion as a function of radius, calculated with $\epsilon=0.01$. The red and blue lines correspond to the calculations at $t=t_1$ and $t=t_2$, respectively.  The size of the MAD is $R_{\rm m}=1.17R_{\rm ISCO}$, when $t=t_2$.}
	\label{b_r}
\end{figure}

\renewcommand{\thefigure}{{S5}}
\begin{figure}
	\centering
	\includegraphics[width=\linewidth]{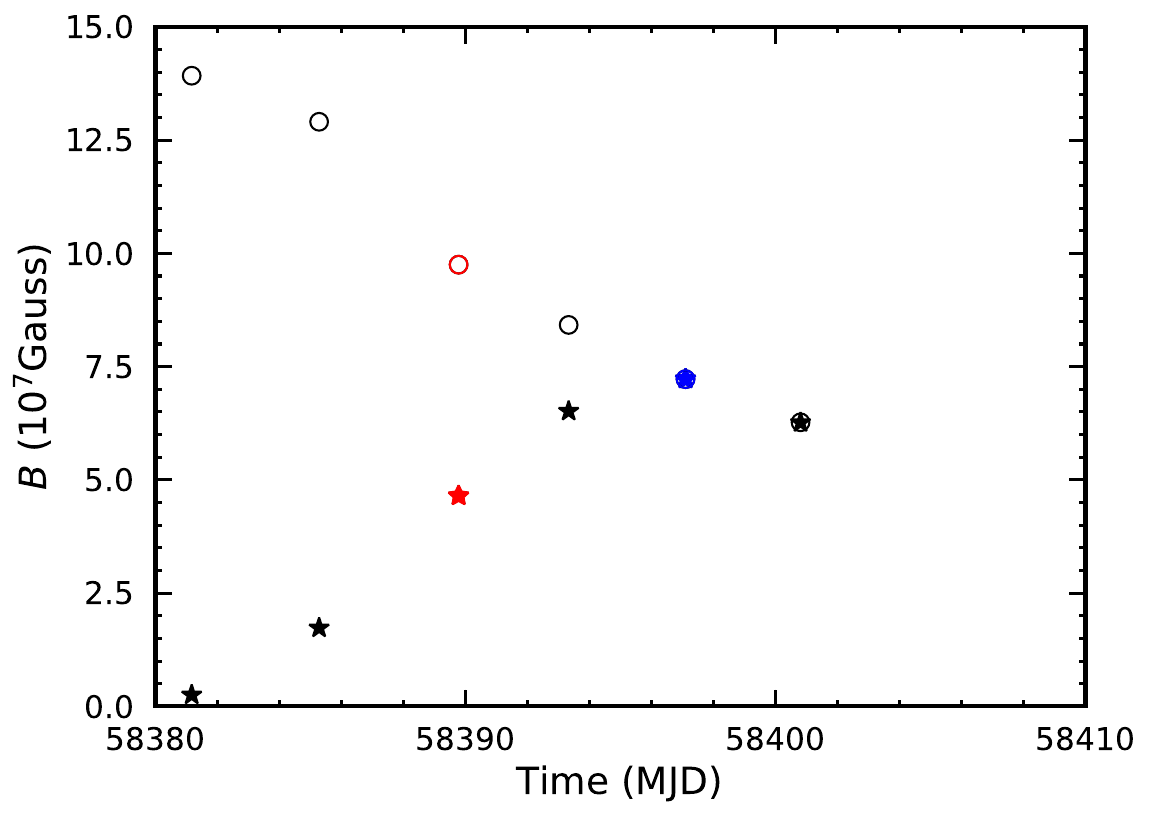}
	\caption{The strength of the magnetic field of the ADAF at $R_{\rm ISCO}$ varies as a function of time. The star points indicate the simulated magnetic strength advected by the ADAF, and the circle points are the MAD criterion. The red and blue colors represent the epochs corresponding to the hard X-ray peak and radio peak, at $t=t_1$ and $t=t_2$, respectively. The black points correspond to other epochs.}
	\label{b_t}
\end{figure}

\renewcommand{\thefigure}{{S6}}
\begin{figure}
	\centering
	\includegraphics[width=\linewidth]{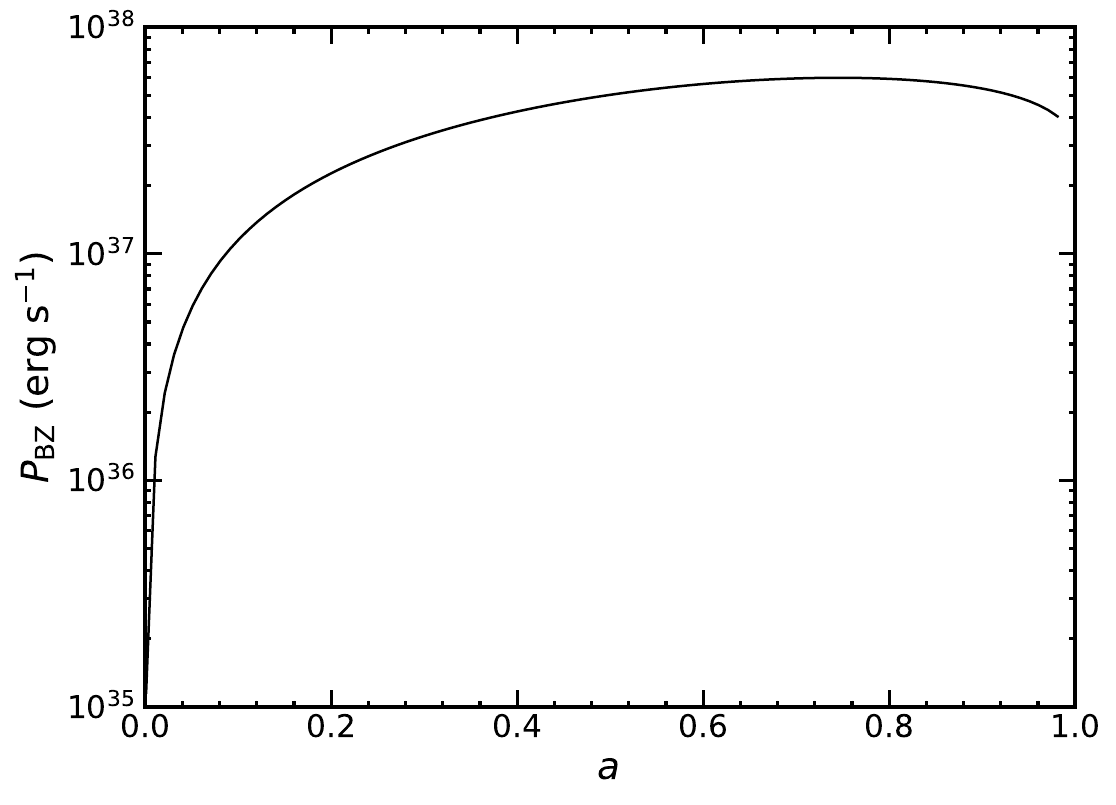}
	\caption{Calculated BZ power, $P_{\rm BZ}$, as a function of the BH spin parameter $a$ for a BH with 8.5 solar mass and the field strength at the horizon {$B_{\rm h}=7\times 10^7$}~Gauss is adopted.}
	\label{jet_power}
\end{figure}

\renewcommand{\thefigure}{{S7}}
\begin{figure}
\centering
\includegraphics[width=\linewidth]{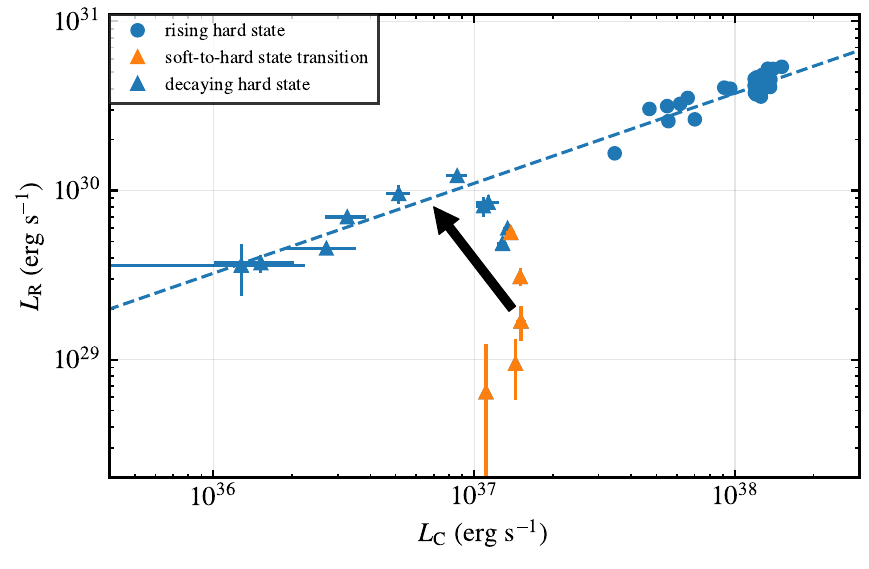} 
\caption{
The correlation between radio and X-ray Comptonization luminosity. The blue dots, blue triangles, and orange triangles represent the data from rising hard state, decaying hard state, and soft-to-hard state transition periods, respectively. The blue dashed line is the best-fitting result of the rising hard state data (blue dots) $L_\mathrm{R} \propto L_\mathrm{X}^{0.53}$. The arrow indicates the evolution of this source in the plane of $L_{\rm R}$-$L_{\rm C}$, during the soft-to-hard state transition. The uncertainties in $L_\mathrm{C}$ and $L_{V}$ are quoted at the 1-$\sigma$ level.
\label{xr}}
\end{figure} 

\renewcommand{\thefigure}{{S8}}
\begin{figure}
\centering
\includegraphics[width=\linewidth]{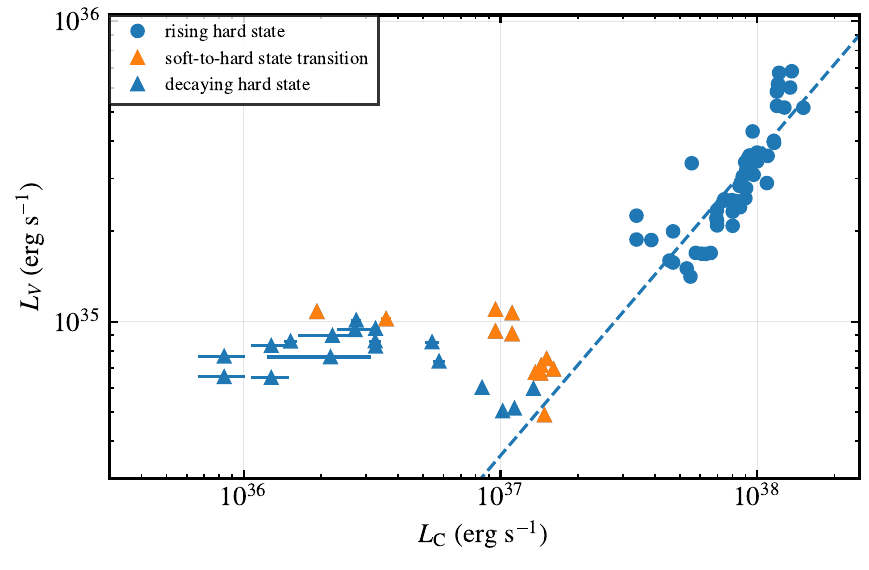} 
\caption{
The correlation between the optical $V$-band and X-ray Comptonization luminosity. The symbols here are the same as the ones in Fig. \ref{xr}. The blue dashed line is the best-fitting result of the rising hard state data (blue dots) $L_V \propto L_\mathrm{C}^{1.00}$. The uncertainties in $L_\mathrm{C}$ and $L_{V}$ are quoted at the 1-$\sigma$ level. 
\label{xv}}
\end{figure}

\clearpage

\noindent {Caption for Movie S1.}\\
 Formation of a MAD in MAXI J1820+070. (A) Animated version of Fig. \ref{schematic_MAD}. (B) The hard X-ray flare arising from the ADAF, the data points in which are estimated from the spectral fits of \textit{Insight}-HXMT observation \cite{methods}, is attributed to the decrease of the mass accretion rate (panel D) and the increase of the ADAF radial extent (panel A). (C) The radio flare originating from the jet, which is observed by AMI-LA, arises from the magnetic field being transported and amplified by the expanding ADAF (panel A). (D) The exponential decay of Eddington-scaled mass accretion rate during the flare, which is predicted based on the disk luminosity and the LE light curve \cite{methods}.



\end{document}